\documentclass[a4paper,12pt,titlepage]{article}
\usepackage{graphicx}
\usepackage{latexsym}
\usepackage[square,numbers,sort]{natbib}

\begin{document}
\ \\[8cm]
\noindent
WHAT DOES THE JOSEPHSON EFFECT TELL US\\
ABOUT THE SUPERCONDUCTING STATE OF THE CUPRATES?\\ \\
\noindent R.~Hlubina$^1$, M.~Grajcar$^1$, and E.~Il'ichev$^2$\\ \\
$^1$Department of Solid State Physics, Comenius University,
SK-84248 Bratislava, Slovakia\\
$^2$ Institute for Physical High
Technology, P.O. Box 100239, D-07702 Jena, Germany

\vspace{1cm}
\noindent
1. INTRODUCTION

\vspace{0.5cm}
\noindent
The defining property of a superconductor is the stiffness of its
phase. In conventional low-$T_c$ superconductors this stiffness is so
large that in order to directly observe its finite value one has to
create a region where it is locally suppressed, the so-called
superconducting weak link.  The resulting weak link region gives rise
to a variety of Josephson effects (for a review, see
\cite{Likharev79}), in which the superconducting phase becomes a
measurable quantity.

In the cuprate superconductors the sensitivity of the Josephson effect
to the phase has played a decisive role in establishing a
phase-sensitive test of the (now well established) unconventional
$d$-wave symmetry of their pairing state (for a review, see
\cite{Tsuei00}).

The purpose of this paper which is based on the results of our
measurements of the current-phase relation in the cuprates
\cite{Ilichev99,Ilichev01,Komissinski02,Grajcar02} is to argue
that besides providing a phase-sensitive test of the pairing symmetry,
a quantitative analysis of the Josephson effect in the cuprates
provides new insights into the microscopic physics of the cuprates, in
particular into the reduced phase stiffness of the cuprates.

Our experiments were carried out on two types of Josephson junctions:
[001] tilt grain boundary junctions
\cite{Ilichev99,Ilichev01,Grajcar02} and $c$-axis junctions between a
cuprate and a low-$T_c$ superconductor \cite{Komissinski02}. Let us
start with a brief discussion of these two types of junctions.
                               
1. Grain boundary junctions have been studied since the very beginning
of the cuprate research (for a review see \cite{Hilgenkamp02}). In
particular, it has been realized very early on that grain boundaries
are ${\it the}$ current limiting element of polycrystalline materials
and as such are to be avoided in large current applications. However,
after having been used in the phase sensitive experiments
\cite{Tsuei00}, the situation changed completely when it was realized
that cuprate grain boundary junctions might form the basis of a
completely new type of superconducting electronics, based on the use
of the so-called $\pi$-junctions.

\begin{figure}[ht]
\centerline{\includegraphics[width=10cm]{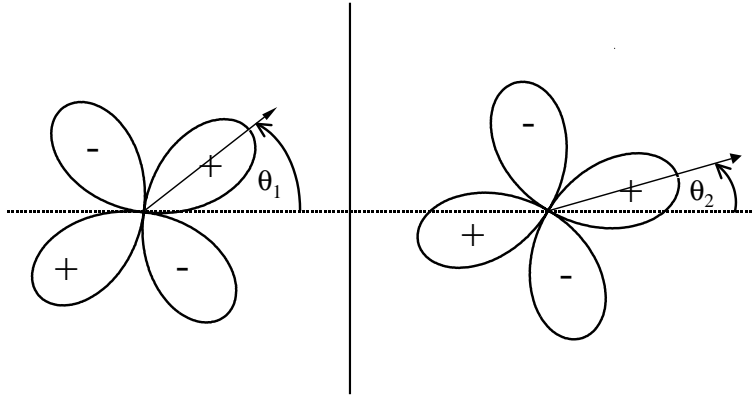}}
\caption{Schematic drawing of a [001] tilt grain boundary junction.}
\label{Fig:Angles_def}
\end{figure}

In an idealized picture, a grain boundary junction can be thought of
as a planar interface of two grains. The junction is characterized by
two angles $\theta_1$ and $\theta_2$ between the normal to the
interface and the crystallographic axes in the grains, see
Fig.~\ref{Fig:Angles_def}.  

It is an experimental fact that the junction transparency ${\cal D}$
depends predominantly on the misorientation angle
$\theta=\theta_2-\theta_1$.  The microstructure of the grain boundary
is quite complicated and at the moment there exists no generally
accepted picture of it.  In this paper we adopt the point of view
advocated by Hilgenkamp and Mannhart \cite{Hilgenkamp02}, who view the
junction as a locally underdoped (and insulating) region. The width of
this insulating barrier is supposed to be proportional to $\theta$,
therefore leading quite naturally to the experimentally observed
scaling ${\cal D}(\theta)=\exp(-\theta/\theta_0)$ with
$\theta_0\approx 5^\circ$. In this paper we shall be concerned mostly
with grain boundary junctions with misorientation angle
$\theta=45^\circ$ for which ${\cal D}\approx e^{-9}\approx 10^{-4}$.

\begin{figure}[ht]
\centerline{\includegraphics[width=10cm]{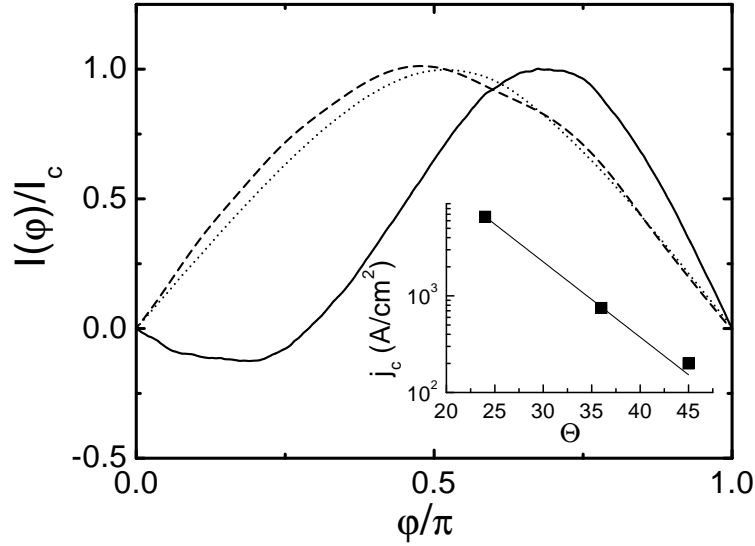}}
\caption{Current-phase relation at $T=15$~K for symmetric GBJJs
with $\theta=24^\circ$ (dotted line), 36$^\circ$ (dashed line) and
45$^\circ$ (solid line). The inset shows the scaling $j_c\propto
\exp(-\theta/\theta_0)$ of the critical current density with
$\theta_0=5.6^\circ$. Taken from \cite{Grajcar02}.} 
\label{Fig:Comp}
\end{figure}

We believe that grain boundary junctions with $\theta>\theta_0$ are to
be described as featureless junctions in the tunnel limit. An
alternative explanation of the very small transmission probability
${\cal D}(\theta)$ could invoke the presence of pinholes in the
barrier, whose density would decrease exponentially with $\theta$. We
believe that measurements \cite{Ilichev99a,Grajcar02} of the relation
between the superconducting current $I$ and the phase difference
$\varphi$ across the junctions with $\theta=24^\circ$ and $36^\circ$
have ruled out this possibility, since the results were fully
consistent with the tunnel limit prediction
$I(\varphi)\propto\sin\varphi$ and no trace of higher harmonics
indicative of small barrier weak links were found, see
Fig.~\ref{Fig:Comp}.

2. The other type of junctions whose current-phase relation we have
studied are the so-called $c$-axis junctions between the cuprates and
low-$T_c$ superconductors.  In this type of junctions the interface is
parallel to the CuO$_2$ planes and the current flows along the $c$
axis of the cuprates. Because of the layered structure of the cuprates
the interlayer forces are presumably much weaker than those within the
layers, and therefore much cleaner and better defined interfaces are
to be expected for this type of junctions, when compared with the
grain boundary junctions.

\vspace{1cm}
\noindent
2. THEORY

\vspace{0.5cm}
\noindent
2.1. GRAIN BOUNDARY JUNCTIONS.  Our main interest in this paper
regards the current-phase relation $I(\varphi)$, which is on general
grounds assumed to be a $2\pi$ periodic function. Moreover, in absence
of magnetic fields we expect that $I(-\varphi)=-I(\varphi)$ and
therefore we can write
\cite{Likharev79}
\begin{equation}
I(\varphi)=I_1\sin\varphi+I_2\sin 2\varphi+\ldots.
\label{eq:Fourier}
\end{equation} 
The $d$-wave symmetry of the pairing state in the cuprates requires
that odd harmonics $I_1,I_3,\ldots$ change sign with
$\theta_i\rightarrow
\theta_i+\pi/2$. If we keep only the lowest-order angular harmonics,
we can therefore write \cite{Walker96}
\begin{equation}  
I_1=I_c\cos 2\theta_1\cos 2\theta_2+  
I_s\sin 2\theta_1\sin 2\theta_2.  
\label{eq:Walker_flat}  
\end{equation}  
The coefficients $I_c,I_s$ are functions of the barrier strength,
temperature $T$, etc.  In particular, if the tunneling is allowed only
for a narrow range of impact angles around normal incidence, only the
$I_c$ term survives. This situation has been considered in the early
paper by Sigrist and Rice \cite{Sigrist92}, who suggested to use the
first term of Eq.~\ref{eq:Walker_flat} as a simple approximate formula
which respects the symmetries of the Ambegaokar-Baratoff like
expression (at $T=0$) for the critical current,
\begin{equation}
{\pi\Delta_k\Delta_q\over eR_N(|\Delta_k|+|\Delta_q|)}
K\left({|\Delta_k-\Delta_q|\over |\Delta_k|+|\Delta_q|}
\right) \rightarrow I_c\cos 2\theta_1\cos 2\theta_2,
\label{eq:ambegaokar}
\end{equation}
where $R_N$ is the resistance of the junction in the normal state,
$\Delta_k$, $\Delta_q$ are gap functions for the case of normal
incidence in the two superconducting electrodes, and $K$ is the
complete elliptic integral.

On the other hand, the even harmonics $I_2,I_4,\ldots$ are not forced
by symmetry to change sign and therefore we neglect their weak angular
dependence (except for the strong dependence on the barrier
transparency ${\cal D}(\theta)$).  This then implies (as first noted
in \cite{Tanaka94}) that for the so-called asymmetric $45^\circ$ grain
boundary junctions (i.e. $\theta_1=45^\circ$, $\theta_2=0^\circ$), the
first harmonic $I_1$ vanishes by symmetry and the current-phase
relation can be approximated by $I(\varphi)=I_2\sin 2\varphi$.

In real life the junction interface meanders around its mean (nominal)
direction. In that case we wish to interpret the function
$I_1(\theta_1,\theta_2)$ as a relation between the local critical
current density and the local interface geometry.  Structural studies
of grain boundaries \cite{Mannhart96} show that the interface is
typically faceted with a typical size of a facet $\approx 100$~nm.
Since this length scale is much larger than both relevant electronic
length scales, the Fermi wavelength and the coherence length, we
believe that the macroscopic first harmonic can be estimated by a
simple averaging of Eq.~\ref{eq:Walker_flat} along the interface.
Making use of such a procedure, if the local junction geometry is
described by $\theta_i+\chi(x)$, $I_1$ of a rough interface is given
by \cite{Grajcar02}
\begin{equation}
I_1={1\over 2}(I_c+I_s)\cos 2\theta
+{x\over 2}(I_c-I_s)\cos 2(\theta_1+\theta_2),
\label{eq:Walker_rough}
\end{equation}
where $x=\langle\cos 4\chi\rangle$ is an interface roughness parameter
and $\langle\ldots\rangle$ denotes an average along the interface.
Note that in the completely rough limit $x=0$, $I_1$ depends only on
the misorientation angle $\theta$ and not on the individual angles
$\theta_i$ \cite{Tsuei00}, since in that limit the notion of the
direction of the interface is meaningless.

On the other hand, since $I_2$ is not forced by symmetry to depend
strongly on $\theta_i$ (except for the dependence through the barrier
transmission on the misorientation angle $\theta$), we expect that
$I_2$ will exhibit only a weak dependence on the surface roughness
$x$.

Summarizing the above symmetry arguments, it follows that $d$-wave
symmetry and sufficient interface roughness (small $x$) imply enhanced
$I_2/I_1$ ratios for all $45^\circ$ junctions.

However, while very robust, the symmetry arguments do not provide
estimates of the ratio $I_2/I_1$.  In what follows we attempt a more
quantitative analysis of the Josephson effect. The increased level of
detail has its prize, however: we have to introduce a microscopic
model of the junction.  Two such models have been discussed in the
literature: a perfectly flat interface which leads to the presence of
midgap states in the case of an impenetrable barrier
\cite{Hu94}, and a model emphasizing the roughness of the interface
and the subsequent development of a nontrivial pattern of currents
flowing along the interface \cite{Millis94}.

\vspace{0.5cm}
\noindent
2.1.1. IDEALLY FLAT JUNCTIONS.  Let us consider first the `flat'
scenario. The key role in this approach is played by the concept of
the so-called midgap states, which should form at surfaces of
anomalous superconductors. Theory predicts that the number of midgap
states should be maximal for (110)-like surfaces of $d$-wave
superconductors and no such states should form at (100) and (001)-like
surfaces \cite{Hu94}. The presence (and the nontrivial surface
orientation dependence) of such midgap states in the cuprates is well
documented by now by STM spectroscopy \cite{Wei98} and by grain
boundary tunneling spectroscopy \cite{Alff98}.  It has been also
suggested that such bound states are at the origin of the observed
nonmonotonic temperature dependence of the penetration depth
\cite{Walter98}.

Let us turn now to the study of the influence of the midgap states on
the Josephson effect in the cuprates, following the early proposal of
Tanaka and Kashiwaya \cite{Tanaka96} (for reviews, see
\cite{Kashiwaya00,Lofwander01}).  Within the simplest description, we
consider a circular Fermi line, specular reflection at the interface,
and a constant order parameter in the superconducting grains. Because
of the translational symmetry along the junction, the scattering
problem factorizes into a set of independent problems for each given
impact angle.

The energy of the Andreev bound states is governed by the probability
$D(\theta,\vartheta)$ of transmission for impact angle $\vartheta$.
The function $D(\theta,\vartheta)$ depends on the details of the
barrier, such as its width and height, none of which are known
reliably in the present context. The only direct experimental measure
of the barrier transmission is the resistance per square of the
junction in the normal state, $R_\Box=(h/2e^2)(\pi d/k_F{\cal D})$,
which provides us with one moment of the function
$D(\theta,\vartheta)$, ${\cal
D}(\theta)=\int_0^{\pi/2}d\vartheta\cos\vartheta D(\theta,\vartheta)$.

\begin{figure}[ht]
\centerline{\includegraphics[width=10cm]{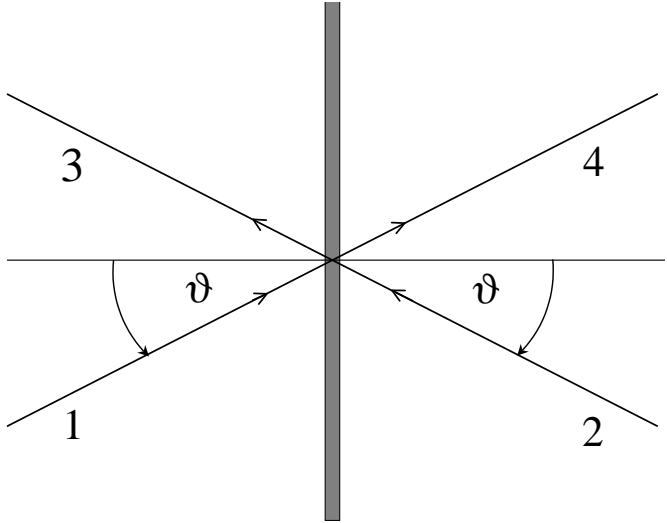}}
\caption{Trajectories of the particles for specular reflection on
a flat junction interface. The impact angle $\vartheta$ is the
same for both sides of the junction.} 
\label{Fig:Angles}
\end{figure}

Let us calculate the energy of the Andreev bound state corresponding
to the incoming (outgoing) trajectories in the left and right grains
denoted 1 and 2 (3 and 4), respectively (see Fig.~\ref{Fig:Angles}).
To this end, let us denote the order parameter on the trajectory $i$
as $\Delta_i=|\Delta_i|S_i$, where $S_i=\pm 1$ is the sign.  The
qualitative character of the bound state depends on the relative value
of the signs $S_i$.

Before proceeding, let us recall that for an impenetrable barrier with
$D\rightarrow 0$, a midgap state is formed in the left (right) grain
for $S_1=-S_3$ ($S_2=-S_4$), i.e. when the sign of $\Delta$ changes
under reflection at the interface \cite{Hu94}.  In the tunnel limit,
the energy of the Andreev bound state can belong to one of the
following three classes:

(i) The signs of $\Delta$ change under reflection on both sides of the
junction, $S_1=-S_3$ and $S_2=-S_4$. In that case the two midgap
states on both sides of the junction are split by the finite value of
$D$ \cite{Barash00}:
\begin{equation}
\varepsilon(\theta,\vartheta,\varphi)=\pm
\sqrt{{4D(\theta,\vartheta)|\Delta_1\Delta_2\Delta_3\Delta_4|}\over
{(|\Delta_1|+|\Delta_3|)(|\Delta_2|+|\Delta_4|)}}
\sqrt{1-S_1S_2\cos\varphi\over 2}.
\label{eq:en_case1}
\end{equation}
Note that $\varepsilon\propto \sqrt{D}$, as is usual for degenerate
perturbation theory.

(ii) The midgap state is formed only on one side of the junction,
the left one for definiteness ($S_1=-S_3$, $S_2=S_4$). In that case 
there is only one anomalous Andreev bound state, \cite{Barash00}:
\begin{equation}
\varepsilon(\theta,\vartheta,\varphi)=
D(\theta,\vartheta)
{|\Delta_1\Delta_3|\over{|\Delta_1|+|\Delta_3|}}
S_1S_2\sin\varphi.
\end{equation}

(iii) There are no midgap states on both sides of the junction, i.e.
$S_1=S_3$ and $S_2=S_4$. In this case there are no amalous Andreev
bound states and the junction resembles qualitatively that between two
$s$-wave superconductors. Note that this type of processes is always
realized for $\vartheta=0^\circ$.

Once the quasiparticle energies are known, the free energy due to
Andreev levels per unit area of the junction reads
\begin{equation}
F_\Box^A(\varphi)=-{k_FT\over 2\pi d}\sum_n\int_{-\pi/2}^{\pi/2}
d\vartheta\cos\vartheta
\ln\left[1+\exp\left(-{\varepsilon_n(\vartheta)\over T}\right)\right],
\label{eq:F_andreev}
\end{equation}
where $d$ is the average distance between the CuO$_2$ planes and the
index $n$ numerates all $\varphi$-dependent energy levels for the
impact angle $\vartheta$.  According to general principles, the
current-phase relation $I(\varphi)$ can be determined from the
$\varphi$-dependence of the total junction free energy,
$I=(2\pi/\Phi_0)\partial F/\partial\varphi$ and the density of
supercurrent across the junction reads
$j(\varphi)=(2\pi/\Phi_0)\partial F_\Box/\partial\varphi$.  We
emphasize that it is the total $\varphi$-dependent free energy density
$F_\Box(\varphi)$ (and not only the contribution of the Andreev bound
states $F_\Box^A$) which determines $j(\varphi)$.

Before proceeding we should like to point out that, in general, a
finite phase difference $\varphi$ across the junction leads to a
nonzero current along the junction \cite{Huck97}.  This current
generates a magnetic field which has to be screened by the Meissner
currents in the superconducting electrodes \cite{Sigrist98}, thus
leading to an additional term in the free energy, $F_\Box^M(\varphi)$.

The spontaneously generated current along the interface can be
calculated as follows.  Since all four momenta involved in interface
scattering with impact angle $\vartheta$ (see Fig.~\ref{Fig:Angles})
have the same momentum $k_y$ along the junction, the $y$ component of
the current density carried by the Andreev bound state with energy
$\varepsilon(\vartheta)$ is
$$
j_y(\vartheta,x)={e\hbar k_y\over m}
\left(\left|u(x,\vartheta)\right|^2+|v(x,\vartheta)|^2\right),
$$
where $u(x)$ and $v(x)$ are the two components of the Andreev level
wavefunction. Since $u(x),v(x)$ decay exponentially on the distance
$\sim\xi$ from the interface (where $\xi\sim \hbar v_F/\Delta$ is the
coherence length), also the net current along the interface 
\begin{equation}
j_y(x)={k_F L\over 2\pi}\sum_n\int_{-\pi/2}^{\pi/2}
d\vartheta\cos\vartheta 
j_y(\vartheta,x)f\left(\varepsilon_n(\vartheta)\right)
\label{eq:j_y}
\end{equation}
is localized within $\xi$ from the interface.  From Eq.~\ref{eq:j_y}
it follows that the linear density of the total current flowing along
the grain boundary per CuO$_2$ layer, $J_0=\int_{-\infty}^\infty
j_y(x)$, is
\begin{equation}
J_0={e\varepsilon_F\over \hbar d}\sum_n\int_{-\pi/2}^{\pi/2}
{d\vartheta\over 2\pi}
\sin 2\vartheta f(\varepsilon_n(\vartheta)),
\label{eq:J_0}
\end{equation}
where $\varepsilon_F$ is the Fermi energy.  In deriving
Eq.~\ref{eq:J_0} we have made use of the fact that the Andreev level
wavefunctions are normalized, $\int_{-\infty}^\infty dx
(|u(x)|^2+|v(x)|^2)=(Ld)^{-1}$.

\begin{figure}[ht]
\centerline{\includegraphics[width=10cm]{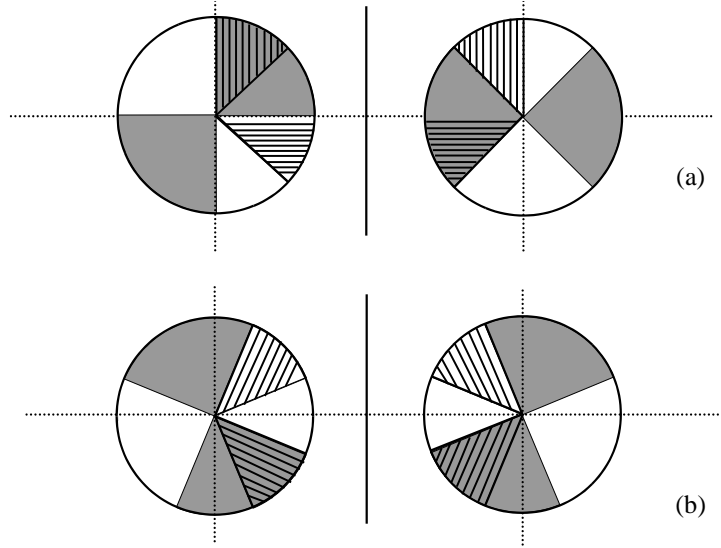}}
\caption{Schematic drawings of asymmetric (a) and symmetric (b)
$45^\circ$ grain boundary junctions. Shaded regions correspond to a
positive order parameter. For asymmetric junctions, impact angles for
which $S_1S_2=-1$ are denoted by the hatched regions. For symmetric
junctions the hatched regions correspond to trajectories with midgap
states on both sides of the junction.}
\label{Fig:Circles}
\end{figure}

{\it Asymmetric $45^\circ$ junctions.} In what follows, let us apply
the above formalism to the two special cases of interest to us, namely
the asymmetric and symmetric $45^\circ$ junctions.  For asymmetric
junctions, the trajectories for all impact angles are of type (ii),
see Fig.~\ref{Fig:Circles}. Since in this case
$\Delta_1(\vartheta)=-\Delta_3(\vartheta)=\Delta\sin 2\vartheta$ and
$\Delta_2(\vartheta)=\Delta_4(\vartheta)=\Delta\cos 2\vartheta$, the
energy of the Andreev levels is
$$
\varepsilon(\vartheta)={\Delta\over 2}D(\vartheta)\:
|\sin 2\vartheta|\:{\rm sign}(\sin 4\vartheta)\sin\varphi.
$$
For the sake of simplicity we consider a simple model for the barrier
transmission function, $D(\vartheta)=D(0)$ for
$|\vartheta|<\vartheta_0<\pi/4$ and $D(\pi/4)\ll D(0)$ otherwise. In
this case ${\cal D}\approx D(0)\sin\vartheta_0$ and the typical energy
of the bound states $\sim {\cal D}\Delta$ sets the energy scale of the
problem. In the two extreme limits $T\ll{\cal D}\Delta$ and $T\gg
{\cal D}\Delta$, respectively, we find
\begin{eqnarray}
F_\Box^A&=&
-{\vartheta_0\over 4\pi}{k_F{\cal D}\Delta\over d}
|\sin\varphi|,
\nonumber
\\
F_\Box^A&=&{\vartheta_0\over 48\pi}
{k_F{\cal D}^2\Delta^2\over Td}
\cos 2\varphi.
\label{eq:FA_asym}
\end{eqnarray}
From Eq.~\ref{eq:J_0} it follows that $J_0$ depends only on the
population of the Andreev levels. Let us for definiteness consider the
case $\varphi>0$. Depending on the impact angle $\vartheta$, the
trajectories can be classified into two distinct groups: The hatched
regions in Fig.~\ref{Fig:Circles} correspond to $S_1S_2=-1$, whereas
for the remaining trajectories $S_1S_2=1$. Thus the energy of the
Andreev levels in the hatched (not hatched) regions is negative
(positive).  Therefore at $T=0$ only the Andreev levels in the hatched
regions are occupied and from Eq.~\ref{eq:J_0} we find $J_0=0$. This
effect has been called superscreening in the literature \cite{Amin01}.
At finite temperatures $T\gg D(\pi/4)\Delta$ an explicit calculation
leads to a finite current along the junction,
$$
J_0=-{\vartheta_0^2\over 4\pi}
{ek_Fv_F\over d}
{{\cal D} \Delta\sin\varphi\over
3T+{\cal D} \Delta |\sin\varphi|}.
$$

The surface currents due to the Andreev bound states generate magnetic
fields which have to be screened by the Meissner currents in the
superconducting grains.  As pointed out in \cite{Lofwander00}, this
leads to an increase of the total junction energy, whose magnitude we
now estimate in the simple case of a two dimensional interface (in the
$y$-$z$ plane) between 3D superconducting grains. Of course, most
experiments are carried out on thin films and thus the 3D geometry is
not fully appropriate. However, since the film thicknesses are
typically $\sim 1000$ \AA\ which is comparable to the bulk penetration
depth, our estimates should remain qualitatively correct.

\begin{figure}[ht]
\centerline{\includegraphics[width=15cm]{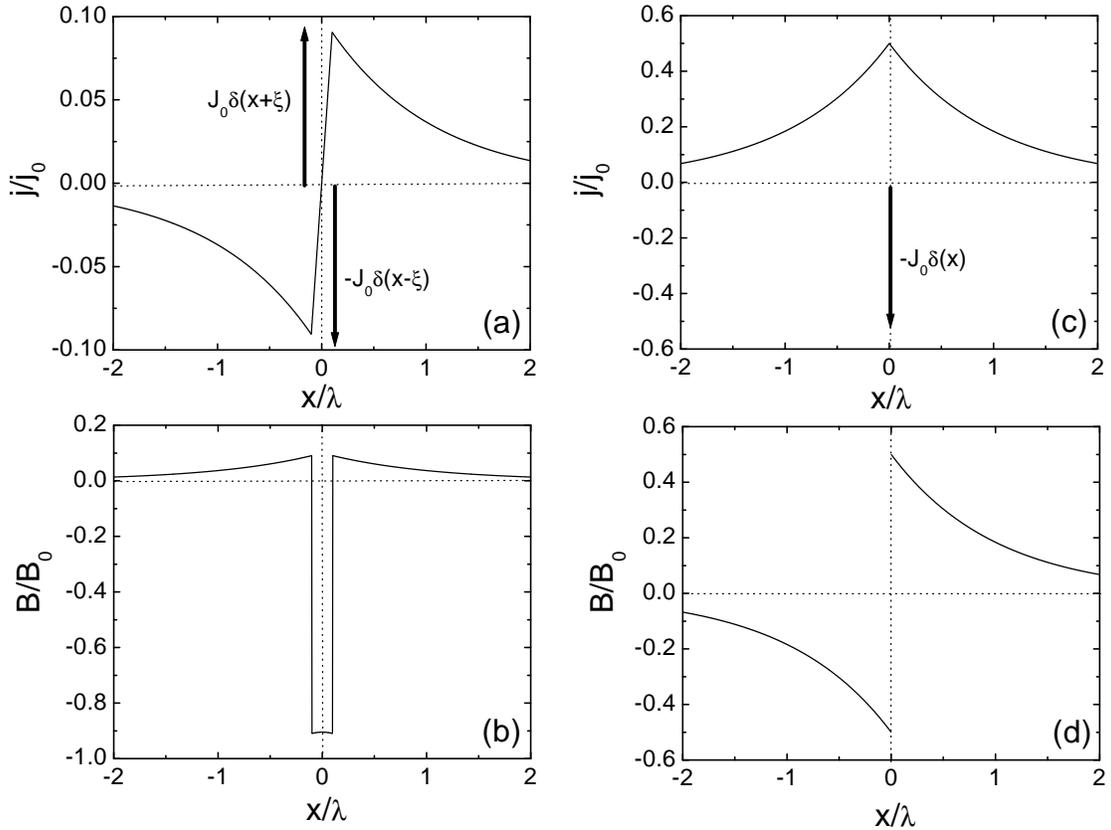}}
\caption{The distribution of supercurrent density (a) in units of
$j_0=J_0/\lambda$ and magnetic field (b) in units of $B_0=\mu_0J_0$
for symmetric $45^\circ$ grain boundary junctions. (c,d): the same for
asymmetric $45^\circ$ junctions. In all figures we have used
$\xi/\lambda=0.1$.}
\label{Fig:Mag}
\end{figure}

Since the current due to Andreev bound states flows on scale $\xi$, we
model its distribution as ${\bf j}=(0,j,0)$ with $j(x)=J_0\delta(x)$.
Within the London theory (assuming bulk penetration depth $\lambda$),
the distribution of the Meissner current and of the magnetic field can
be readily calculated and the results are shown in
Fig.~\ref{Fig:Mag}. The sum of the energy of the magnetic field and of
the kinetic energy of the superflow (per unit area of the interface)
is easily seen to be $F_\Box^M=\mu_0 J_0^2\lambda/4$.

Therefore the $\varphi$-dependent part of the total free energy
density $F_\Box(\varphi)=F_\Box^A(\varphi)+F_\Box^M(\varphi)$ for
$T\ll {\cal D}\Delta$ and $T\gg {\cal D}\Delta$ reads
\begin{eqnarray}
F_\Box(\varphi)&=&{\vartheta_0\over 4\pi}{k_F\over d}\Delta
\left[-{\cal D} |\sin\varphi|+
{\vartheta_0^3\over 8}{\xi\over\lambda}
\left(
{\sin\varphi\over |\sin\varphi|+\alpha}
\right)^2
\right],
\nonumber\\
F_\Box(\varphi)&=&{\vartheta_0\over 48\pi}{k_F\over d}
{{\cal D}^2\Delta^2\over T}
\left[1-{\vartheta_0^3\over 12}{\hbar v_F/T\over\lambda}
\right]\cos 2\varphi,
\label{eq:F_asym}
\end{eqnarray}
respectively, where $\xi=\hbar v_F/\Delta$ and $\alpha=3T/{\cal
D}\Delta$. Several points regarding Eq.~\ref{eq:F_asym} are to be
stressed. First, in the low temperature limit, our result agrees
qualitatively with \cite{Lofwander00} in that the free energy is
minimized for $\varphi=0$ or $\varphi=\pi$, if ${\cal D}\ll
\vartheta_0^3\xi/\lambda$ (i.e. the bound state energy gain is smaller
than the Meissner energy loss).  In the opposite limit when the bound
state energy dominates $F_\Box$, the free energy is minimized for
$\varphi=\pm\pi/2$.

\begin{figure}[ht]
\centerline{\includegraphics[width=9cm]{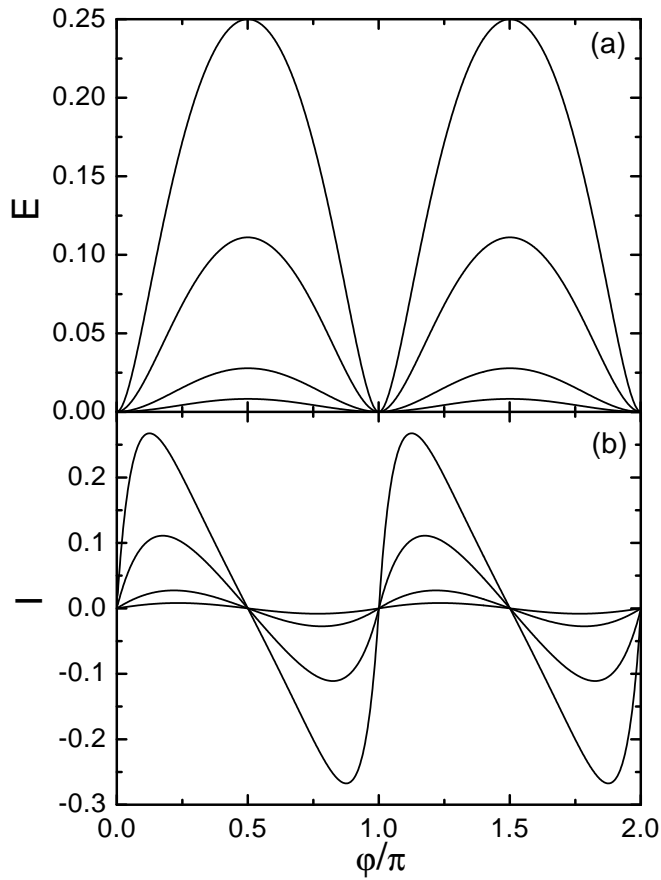}}
\caption{(a) Meissner energy $F^M$ of an asymmetric $45^\circ$
junction (in arbitrary units) as a function of the phase difference
$\varphi$ across the junction. (b) Superconducting current
$I=(2\pi/\Phi_0)\partial F^M/\partial\varphi$ (in arbitrary units) as
a function of $\varphi$. In both figures $\alpha=1,2,5,10$ from
top to bottom.}
\label{Fig:Ipsteep}
\end{figure}

From the normal state resistivity of $45^\circ$ junctions we estimate
${\cal D}\sim 10^{-4}$. If we take $\xi\approx 30$~\AA\ and
$\lambda\approx 1500$~\AA\ relevant for the cuprates, we estimate that
for not too strongly angle dependent tunneling, the Meissner energy
should dominate $F_\Box$ at low temperatures. We should like to point
out that by no means this implies trivial physics. Just on the
contrary, as shown in Fig.~\ref{Fig:Ipsteep}, dominant Meissner energy
leads to a doubled periodicity of $I(\varphi)$. This is qualitatively
similar to the well studied case when $F_\Box^A$ dominates. However,
there are two major qualitative differences with respect to the case
of dominant bound state energy: (i) the (degenerate) ground state
corresponds to a state when no currents flow along the junction, and
(ii) the current-phase relation becomes steep in the minima (not
maxima) of the junction energy, see Fig.~\ref{Fig:Ipsteep}.

At temperatures above the bound state energy ${\cal D}\Delta$
(which is presumably the case down to helium temperatures due to the
small value of ${\cal D}$), we find $F_\Box\propto \cos
2\varphi$ leading to  
\begin{equation}
j(\varphi)R_\Box=-{\pi\vartheta_0\over 12}
{{\cal D}\Delta^2\over eT}
\left[1-{\vartheta_0^3\over 12}{\hbar v_F/T\over\lambda}
\right]\sin 2\varphi.
\label{eq:iphi_asym}
\end{equation}
A pure second harmonic is seen to develop, in complete agreement with
the symmetry analysis. Note that in this temperature range the
Meissner contribution to $I(\varphi)$ becomes negligible. This is a
result of the thermal fluctuations which reduce the surface current
$J_0$ severely.

{\it Symmetric $45^\circ$ junctions,} i.e. junctions with
$\theta_1=-\theta_2=22.5^\circ$.  In this case, trajectories with
impact angles in the hatched regions of Fig.~\ref{Fig:Circles} are of
type (i), whereas the remaining ones are of type (iii). As a function
of the impact angle, the gap functions can be written
$\Delta_1(\vartheta)=\Delta_2(\vartheta)=\Delta\cos(2\vartheta-\pi/4)$
and
$\Delta_3(\vartheta)=\Delta_4(\vartheta)=\Delta\cos(2\vartheta+\pi/4)$.
Again we have to make an assumption about the dependence of the
transparency on the impact angle. We start by considering the case of
a weakly dependent $D(\vartheta)$.  As a rough approximation, we
approximate all Andreev levels of type (iii) by their value for
$\vartheta=0^\circ$ \cite{Barash00}, and all levels of type (i) by the
$\vartheta=45^\circ$ case,
\begin{eqnarray*}
\varepsilon(0,\varphi)&=&\pm 
2^{-1/2}\Delta\sqrt{1-D(0)\sin^2(\varphi/2)},
\\
\varepsilon(\pi/4,\varphi)&=&\pm
2^{-1/2}\Delta\sin(\varphi/2)\sqrt{D(\pi/4)}, 
\end{eqnarray*}
where we assume that $D(\pi/4)<D(0)\ll 1$.  Note that for every impact
angle $\vartheta$ there exist two levels with energies
$\pm\varepsilon(\vartheta)$.  Therefore $\sum_n
f(\varepsilon_n(\vartheta))=1$ and the integral Eq.~\ref{eq:J_0}
vanishes.  In other words, the total supercurrent along the interface
vanishes by symmetry.  This does not mean, however, that no currents
are flowing along the interface.  In order to estimate the Meissner
contribution to the interface free energy $F_\Box^M$, we model the
current distribution generated by the Andreev levels as
$j(x)=J_0[\delta(x+\xi)-\delta(x-\xi)]$ where $\xi$ is the coherence
length and $J_0=\int_0^\infty dx j_y(x)$.  The resulting magnetic
field and the Meissner screening currents can be calculated within the
London theory in a similar way as for the asymmetric junctions (see
Fig.~\ref{Fig:Mag}) and the resulting contribution to the junction
free energy is $F_\Box^M=\mu_0 J_0^2\xi$. Note that $F_\Box^M$ is
reduced with respect to the case of asymmetric junctions (if we assume
the same value of $J_0$ in both cases) by a factor $4\xi/\lambda\ll
1$, since it arises from a higher order moment of the current
distribution $j_y(x)$. Due to this small factor,
$F_\Box^M/F_\Box^A\sim (\xi/\lambda)^2/D$ even if we take the maximal
possible value of $J_0$, $J_0\sim ek_Fv_F/d$. Therefore the Meissner
contribution to the interface free energy will be neglected and the
Josephson current is calculated from
\begin{equation}
j(\varphi)={2\pi\over\Phi_0}{\partial F_\Box^A\over\partial\varphi}=
{k_F\over\Phi_0 d}\sum_n\int_{-\pi/2}^{\pi/2} d\vartheta\cos\vartheta
f(\varepsilon_n(\vartheta)){\partial\varepsilon_n\over\partial\varphi}.
\label{eq:j_andreev}
\end{equation}
This implies that $j(\varphi)=j_n(\varphi)+j_a(\varphi)$, where the
normal and anomalous contributions are, respectively:
\begin{eqnarray}
j_n(\varphi)&=&{\pi\over 8}{k_F\Delta\over\Phi_0d}
{D(0)\sin\varphi\over 2^{3/2}\sqrt{1-D(0)\sin^2(\varphi/2)}}
\tanh{\Delta\sqrt{1-D(0)\sin^2(\varphi/2)}\over 2^{3/2}T},
\nonumber
\\
j_a(\varphi)&=&-{\pi\over 8}{k_F\Delta\over\Phi_0d}
\cos(\varphi/2)\sqrt{D(\pi/4)}\tanh
{\Delta\sin(\varphi/2)\sqrt{D(\pi/4)}\over 2^{3/2}T}.
\label{eq:iphi_sym1}
\end{eqnarray}
At intermediate temperatures $\Delta\sqrt{D(\pi/4)}\ll T\ll \Delta$
the anomalous (normal) Andreev levels are in the high (low)
temperature limit. Therefore Eq.~\ref{eq:iphi_sym1} simplifies
considerably and the first two harmonics of $j(\varphi)$ can be
written as
\begin{eqnarray}
{j_1\over j_L}&=&D(0)-{\Delta\over 2T}D(\pi/4),
\nonumber\\
{j_2\over j_L}&=&-{1\over 8}D(0)^2-
{1\over 24}\left({\Delta\over 2T}\right)^3D(\pi/4)^2,
\label{eq:iphi_sym2}
\end{eqnarray}
where $j_L=(\pi/16\sqrt{2})k_F\Delta/\Phi_0 d$ is comparable with the
bulk depairing current density. Thus, for weakly angle-dependent
tunneling and above the energy of anomalous Andreev levels, theory
predicts a sign change of the first harmonic at a temperature
$T^\ast=\Delta D(\pi/4)/2D(0)$ and a second harmonic monotonically
increasing with decreasing temperature.

In the opposite limit of a peaked barrier transparency
$D(\vartheta)=D(0)\exp(-\vartheta/\vartheta_0)$, the junction is
equivalent to a symmetric junction between $s$-wave superconductors
with order parameters $\Delta/\sqrt{2}$. Therefore the current-phase
relation in the $T\rightarrow 0$ and $T\rightarrow T_c$ limits,
respectively, reads
\begin{eqnarray}
j(\varphi)R_\Box&=&{\pi\over 2\sqrt{2}}{\Delta\over e}
\left[\sin\varphi-{D(0)\over 16}\sin 2\varphi\right],
\nonumber\\
j(\varphi)R_\Box&=&{\pi\over 8}{\Delta^2\over eT}
\left[\sin\varphi-{D(0)\over 48}\left({\Delta\over 2T}\right)^2
\sin 2\varphi\right].
\label{eq:iphi_sym3}
\end{eqnarray}
Unlike Eq.~\ref{eq:iphi_sym2}, the result Eq.~\ref{eq:iphi_sym3}
predicts a monotonically increasing first harmonic with decreasing
temperature. 

\vspace{0.5cm}
\noindent
2.2. $c$-AXIS JUNCTIONS.  Let us briefly discuss the supercurrent in
$c$-axis junctions between YBa$_{2}$Cu$_{3}$O$_{x}$ (YBCO) and
low-$T_c$ superconductors \cite{Komissinski02}.  In YBCO the dominant
component of the superconducting order parameter has $d$-wave symmetry
\cite{Tsuei00}.  However, due to the orthorhombic structure of YBCO, a
finite component with $s$-wave symmetry is admixed to the dominant
$d$-wave order parameter. The in-plane phase sensitive experiments
imply that the $d$-wave component remains coherent through the whole
sample \cite{Tsuei00}, while an elegant $c$-axis tunneling experiment
shows directly that the $s$-wave order parameter component does change
sign across the twin boundary \cite{Kouznetsov97}.

The above picture of the YBCO pairing state is challenged by the
experimental observation of a finite $c$-axis Josephson current
between heavily twinned YBCO and a Pb counterelectrode
\cite{Sun96}. Namely, the contribution of the $s$-wave part of the
YBCO order parameter to the Josephson coupling between a conventional
superconductor (superconductor with a pure $s$-wave symmetry of the
order parameter, for example Pb or Nb) and YBCO should average to zero
for equal abundances of the two types of twins in YBCO. In other
words, the macroscopic pairing symmetry of twinned YBCO samples should
be a pure $d$-wave \cite{Walker96}. Tanaka has shown that a finite
second order Josephson current obtains for a junction between the
$s$-wave and $c$-axis oriented pure $d$-wave superconductors \cite
{Tanaka94}. However, measurements of microwave induced steps at
multiples of $hf/2e$ on the $I$-$V$ curves of Pb/Ag/YBCO tunnel
junctions imply dominant first order tunneling
\cite{Kleiner96}. Therefore the finite $c$-axis Josephson current has to
result from a nonvanishing admixture of the $s$-wave component to the
macroscopic order parameter of YBCO \cite{Walker96}.  Two alternatives
of how this can take place in the junctions based on twinned YBCO have
been discussed in the literature:

(i) Sigrist \textit{et al.} have suggested that the phase of the
$s$-wave component in YBCO does not simply jump from 0 to $\pi $ upon
crossing the twin boundary, but rather changes in a smooth way,
attaining the value of $\pi /2$ right at the twin boundary
\cite{Sigrist96}. The twinned YBCO sample is thus assumed to exhibit a
macroscopic $d+is$ pairing symmetry. A related picture has been
proposed by Haslinger and Joynt, who suggest a $d+is$ surface state of
YBCO \cite{Haslinger00}.

(ii) A difference in the abundances of the two types of twins implies
a $d+s$ symmetry of the macroscopic pairing state
\cite{ODonovan97}. It has been pointed out \cite{Komissinski02} 
that also structural peculiarities of other type (such as a lamellar
structure in a preferred direction) may lead to the $d+s$ macroscopic
pairing symmetry.

\begin{figure}[ht]
\centerline{\includegraphics[width=14cm]{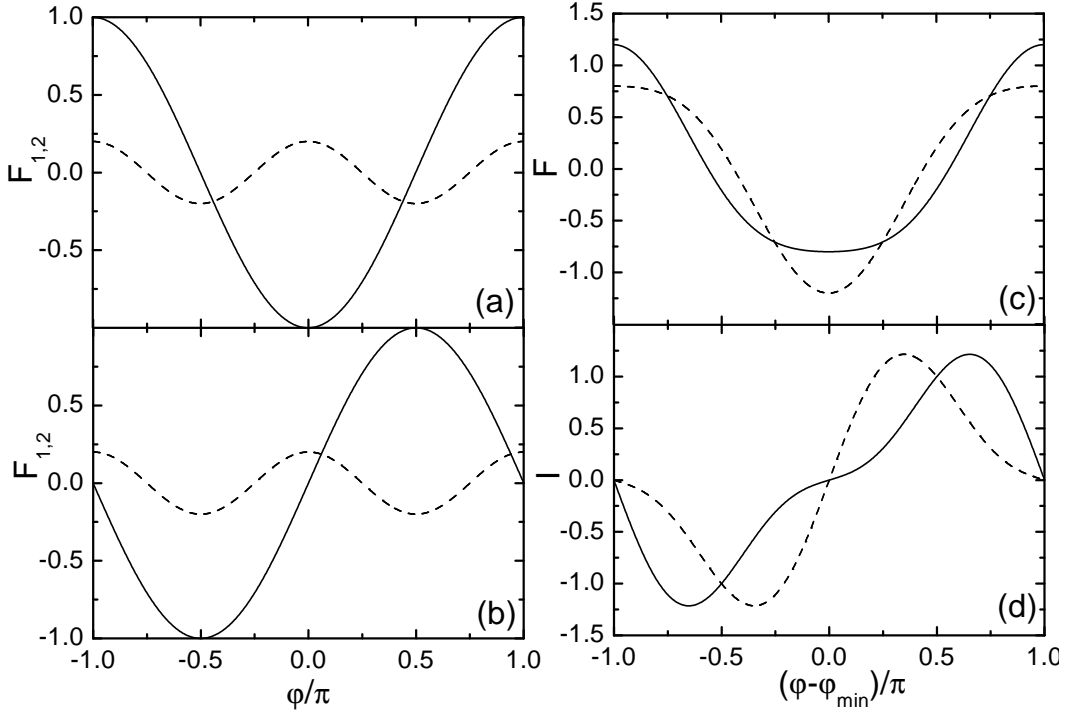}}
\caption{(a,b): Schematic drawing of the contributions from the first 
(solid lines) and second (dashed lines) harmonics to the free energy
of $c$-axis junctions between a conventional superconductor and YBCO
for various macroscopic pairing states in YBCO. (a): $d+s$ pairing,
(b): $d+is$ pairing. (c,d): Total free energy (c) and the
current-phase relation (d) in the $d+s$ case (solid lines) and $d+is$
case (dashed lines).}
\label{Fig:Signs}
\end{figure}

The primary motivation of our experiment \cite{Komissinski02} was to
determine which of the above two scenaria is realized in YBCO. In what
follows we will show that this question can be conveniently answered
by measuring the relative sign of the first two harmonics of
$I(\varphi)$.  Let us consider first the $d+s$ scenario.  As will be
shown explicitly in Eqs.~\ref{eq:j_1}, \ref{eq:j_2}, in this case the
junction free energy can be written
\begin{equation}
F(\varphi)=-F_1\cos\varphi+F_2\cos 2\varphi
\label{eq:sign1}
\end{equation}
with $F_1,F_2>0$. As shown in Fig.~\ref{Fig:Signs}, this choice of
signs corresponds to a free energy which is flat (curved) close to the
minima (maxima). In terms of the current phase relation, this case
corresponds to the usual case when $I(\varphi)$ is steep in the
vicinity of $\pm\pi$, as in the Kulik-Omelyanchuk theory
\cite{Likharev79}. 

The case of $d+is$ pairing is more complicated. In fact, first order
coupling between the conventional superconductor and the $s$-wave
component leads to a term in free energy $-F_1\cos(\varphi_{\rm
conv}-\varphi_s)$, where $\varphi_{\rm conv}$ and $\varphi_s$ are the
phases of the conventional superconductor and of the $s$-wave
component in YBCO and $F_1>0$. On the other hand, the second order
term is the same as in the $d+s$ scenario. Therefore, since
$\varphi=\varphi_{\rm conv}-\varphi_d$ where $\varphi_d$ is the phase
of the $d$-wave component in YBCO, and since
$\varphi_s-\varphi_d=\pi/2$ in the $d+is$ scenario, the junction free
energy can be written
\begin{equation}
F(\varphi)=-F_1\cos(\varphi-\varphi_{\rm min})
-F_2\cos 2(\varphi-\varphi_{\rm min}),
\label{eq:sign2}
\end{equation}
where $\varphi_{\rm min}=\pi/2$ is the equilibrium phase difference
and $F_1,F_2>0$, see Fig.~\ref{Fig:Signs}.  In the experiment it is a
delicate issue to determine the absolute value of $\varphi_{\rm
min}$. Usually $\varphi_{\rm min}$ is set equal to zero in the free
energy minimum and with such conventions the two cases
Eqs.~\ref{eq:sign1},\ref{eq:sign2} lead to a different relative sign
of $I_1$ and $I_2$ for the $d+s$ and $d+is$ scenaria, respectively.

\vspace{0.5cm}
\noindent
2.3. FACETED JUNCTIONS. In situations when the first harmonic $I_1$
vanishes by symmetry (i.e. for asymmetric $45^\circ$ grain boundary
junctions or for $c$-axis tunneling between an $s$-wave and a $d$-wave
superconductor), small local deviations from ideal geometry lead to a
finite local coupling $\delta I_1$ across the junction. In such cases
the junction can be viewed as a parallel series of Josephson junctions
with local critical current whose magnitude and sign fluctuate along
the interface.  It is customary to call the junctions with a positive
(negative) $\delta I_1$ as 0 and $\pi$ junctions, respectively.  Let
us note in passing that very recently it has become possible to
prepare samples with a prescribed pattern of 0 and $\pi$ junctions,
making use of the so-called zigzag junctions \cite{Smilde02}.

\begin{figure}[ht]
\centerline{\includegraphics[width=10cm]{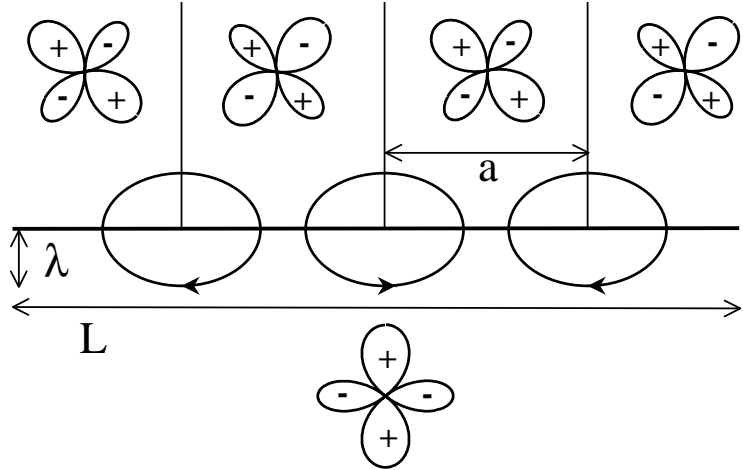}}
\caption{Schematic drawing of an asymmetric $45^\circ$ grain boundary
junction between twinned YBCO superconducting grains. In the upper
grain the sign of the dominant lobe of the order parameter changes
between the neighboring twins, leading to a sequence of 0 and $\pi$
junctions. Also shown are the spontaneously generated currents along
the interface.}
\label{Fig:Facet}
\end{figure}

An example of a series of 0 and $\pi$ junctions is shown in
Fig.~\ref{Fig:Facet}, where the geometry of an asymmetric $45^\circ$
grain boundary between two twinned YBCO samples is shown.  Twinning in
the grain with $\theta_1=0^\circ$ is not shown explicitly, since it is
irrelevant for our argument. Another often considered source of
deviations from ideal interface geometry is associated with the
experimentally well documented meandering of interface
\cite{Mannhart96}.

If the average interface geometry is close to ideal, the number of 0
and $\pi$ junctions is roughly the same.  As noted first by Millis
\cite{Millis94} (for an alternative formulation making use of the 
sine-Gordon equation, see e.g. \cite{Mannhart96,Mints98}), this does
not imply a vanishing Josephson coupling across the interface. The
general idea is as follows: let the average phase difference across
the junction is $\varphi$. The phase difference along the interface is
modulated from this value towards $0$ and $\pi$ in the 0 and $\pi$
junction regions, respectively, by an amount $\sim\chi$.  This leads
to spontaneously generated currents along the interface which are
schematically depicted in Fig.~\ref{Fig:Facet}.  In this way, the
junction gains Josephson energy $\propto\chi$, whereas a Meissner
energy of only $\propto\chi^2$ is lost. The total energy gain is
maximal for $\varphi=\pm\pi/2$ and this explains qualitatively the
development of a second harmonic in such junctions.

\begin{figure}[ht]
\centerline{\includegraphics[width=10cm]{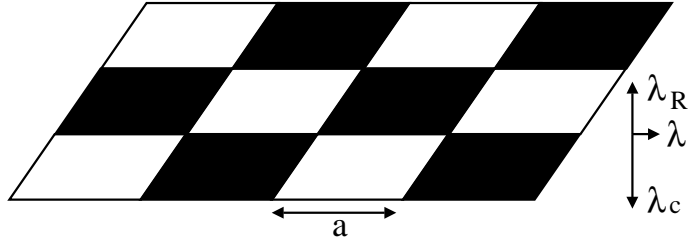}}
\caption{Schematic drawing of a $c$-axis junction between an $s$-wave
superconductor and a twinned YBCO sample. The black and white regions
(of typical size $a$) correspond to $0$ and $\pi$ junctions,
respectively. Also shown are the penetration depths in the $s$-wave
superconductor ($\lambda_R$) and in YBCO ($\lambda,\lambda_c$).}
\label{Fig:Checkboard}
\end{figure}

In what follows, let us discuss the faceted scenario in more detail.
For definiteness, we consider the case of a $c$-axis junction between
an $s$-wave superconductor and a twinned YBCO sample, see
Fig.~\ref{Fig:Checkboard}.  For the sake of simplicity, we model the
randomly twinned structure by a checkerboard distribution of 0 and
$\pi$ junctions with a periodic distribution of the local critical
current density $j_c({\bf x})=j_0 g({\bf x})$, where $g({\bf
x})=\sin(\pi x/a)\sin(\pi y/a)$ and for $a$ we take the typical size
of the twins. Here we have assumed a coordinate system such that the
interface lies in the $xy$ plane. According to \cite{Millis94}, a
spontaneous magnetic field is generated, which we now find by
minimization of the total junction energy as a function of the phase
difference $\varphi$ across the junction. The symmetry of the problem
dictates that the magnetic field reads
\begin{eqnarray}
B_x&=&\sum_n^\prime B_n \exp(-K_n|z|)
\sin(n\pi x/a) \cos(n\pi y/a),
\nonumber
\\
B_y&=&-\sum_n^\prime B_n \exp(-K_n|z|)
\cos(n\pi x/a) \sin(n\pi y/a),
\nonumber
\\
B_z&=&0,
\label{eq:b_millis}
\end{eqnarray}
where the prime restricts the summation to odd values of $n$. Note
that $\nabla\cdot {\bf B}=0$, as it should be. For $z>0$, i.e. in the
conventional superconductor, the ${\bf B}$ field satisfies the
equation $(\nabla^2-\lambda_R^{-2}){\bf B}=0$ if
$K_n^2=\lambda_R^{-2}\left[1+2n^2\left(\pi\lambda_R/a\right)^2\right]$.
On the other hand, in the anisotropic cuprates, ${\bf B}=(B_x,B_y,0)$
is governed by $B_{x,y}=\left[\lambda_c^2(\partial_x^2+\partial_y^2)+
\lambda^2\partial_z^2\right]B_{x,y}$ and therefore for $z<0$ we have 
to take $K_n^2=\lambda^{-2}\left[1+
2n^2\left(\pi\lambda_c/a\right)^2\right]$. From Eq.~\ref{eq:b_millis},
the distribution of Meissner screening currents can be calculated
using $\nabla\times{\bf B}=\mu_0{\bf j}$. The total junction energy
per unit area
corresponding to the magnetic field described by
Eq.~\ref{eq:b_millis} is
\begin{equation}
E_\Box={1\over 4\mu_0}\sum_n^\prime B_n^2\lambda_n
-{\Phi_0j_0\over 2\pi S}\int dS g({\bf x})\cos\theta({\bf x}),
\label{eq:energy_millis}
\end{equation}
where $S$ is the junction area and
$$
\lambda_n=\lambda_R\sqrt{1+2n^2(\pi\lambda_R/a)^2}+
\lambda\sqrt{1+2n^2(\pi\lambda_c/a)^2}.
$$
The local phase difference across the junction $\theta({\bf x})$ can
be determined from the Josephson equations
\begin{eqnarray}
{\partial\theta\over\partial x}&=&
{2\pi\mu_0\over\Phi_0}\left[\lambda^2j_x(z=0_{-})-
\lambda_R^2j_x(z=0_{+})\right]=
{2\pi\over\Phi_0}\sum_n^\prime
B_n\lambda_n\cos{n\pi x\over a}\sin{n\pi y\over a},
\nonumber
\\
{\partial\theta\over\partial y}&=&
{2\pi\mu_0\over\Phi_0}\left[\lambda^2j_y(z=0_{-})-
\lambda_R^2j_y(z=0_{+})\right]=
{2\pi\over\Phi_0}\sum_n^\prime
B_n\lambda_n\sin{n\pi x\over a}\cos{n\pi y\over a}.
\end{eqnarray}
These equations are solved by $\theta({\bf x})=\varphi+\chi({\bf
x})$, where $\varphi$ is the average phase difference and
\begin{equation}
\chi({\bf x})={2a\over\Phi_0}\sum_n^\prime
{B_n\lambda_n\over n}\sin{n\pi x\over a}\sin{n\pi y\over a}
\end{equation}
is the spontaneously generated modulation of $\theta({\bf x})$.

The Fourier components $B_n$ have to be determined by minimization of
the energy Eq.~\ref{eq:energy_millis}. To this end let us assume now
that $|\chi|\ll 1$, which assumption will be justified at the end of
the calculation. The second term in Eq.~\ref{eq:energy_millis}
simplifies in this case to $(2\mu_0)^{-1}B_1B_{\rm
eff}\lambda_1\sin\varphi$, where $B_{\rm eff}=(2\pi)^{-1}\mu_0j_0a$.
Note that Fourier components $B_n$ with $n\geq 3$ raise the energy and
therefore vanish. This means that the junction energy is at the end a
function only of $B_1$, $E_\Box=(\lambda_1/4\mu_0)\left[(B_1+B_{\rm
eff}\sin\varphi)^2-B_{\rm eff}^2\sin^2\varphi\right]$.  The minimum is
reached for $B_1=-B_{\rm eff}\sin\varphi$. The minimized value of the
junction energy density and the resulting current-phase relation read
\begin{eqnarray}
E_\Box(\varphi)&=&-{B_{\rm eff}^2\lambda_1\over 4\mu_0}
\sin^2\varphi,
\nonumber\\
j(\varphi)&=&-{\pi B_{\rm eff}^2\lambda_1\over 2\mu_0\Phi_0}
\sin 2\varphi.
\label{eq:j_millis}
\end{eqnarray}
Note that $j\propto j_0^2\propto{\cal D}^2$. Therefore the second
harmonic scales with the barrier transparency in the same way both
within the flat (for $T\gg{\cal D}\Delta$) and the faceted
scenaria. Finally, let us point out that the result
Eq.~\ref{eq:j_millis} is valid only for $|\chi|\ll 1$, or equivalently
for
\begin{equation}
B_{\rm eff}\ll{\Phi_0\over 2a\lambda_1}.
\label{eq:crit_millis}
\end{equation}

\vspace{0.5cm}
\noindent
2.4. THE JOSEPHSON PRODUCT.  According to standard theory (for
homogeneous featureless barriers), the Josephson product $I_1R_N$
(where $R_N$ is the junction resistance in the normal state) is
independent of the junction area and of the barrier transparency, thus
giving an intrinsic information about the superconducting banks.  The
measured Josephson product of cuprate grain boundary junctions
\cite{Hilgenkamp02} can be well described by
$I_1R_N=\alpha^2(\pi/4)(\Delta/e)\cos 2\theta$, where $\Delta$ is the
maximal superconducting gap.  The angular dependence of $I_1R_N$ is
fully consistent (apart from the $\theta$-independent renormalization
factor $\alpha^2\sim 10^{-1}$) with the BCS prediction
Eq.~\ref{eq:Walker_rough} for rough junctions between $d$-wave
superconductors.  This means that
\begin{equation}
(I_c+I_s)R_N=\alpha^2(\pi/2)(\Delta/e),
\label{eq:Jos_prod} 
\end{equation}
much smaller than the theoretical prediction Eq.~\ref{eq:ambegaokar}
for $I_cR_N$. Note that the interpretation of a small factor
$\alpha^2$ as a result of a nearly complete cancellation of $I_c$ and
$I_s$ is not plausible, since the cancellation would have to occur for
all misorientation angles, whereas the physical origin of the $I_c$
and $I_s$ terms is quite different.

It is worth pointing out
that Eq.~\ref{eq:Jos_prod} applies (with the same $\alpha^2\sim
10^{-1}$) also for the break junctions (in which the misorientation
angle $\theta=0$) \cite{Miyakawa99}. Moreover, in \cite{Miyakawa99} it
has been shown that $\Delta$ is not depressed in the junction region,
thus explicitly demonstrating the breakdown of the BCS prediction for
$I_1R_N$ in the cuprates.

There exists no generally accepted explanation of the small
renormalization factor $\alpha^2$. One of the reasons is that the
microstructure of Josephson junctions is typically quite complicated.
In fact, it is well known that small angle grain boundaries can be
modelled by a sequence of edge dislocations, while at larger
misorientation angles the dislocation cores start to overlap and no
universal picture applies to the structure of the grain boundary.  For
large-angle grain boundaries, Halbritter has proposed
\cite{Halbritter92} that the junction can be thought of as a nearly
impenetrable barrier with randomly placed highly conductive channels
across it. If due to strong Coulomb repulsion only the normal current
(and no supercurrent) is supported by these channels, the small value
of $I_1R_N$ follows quite naturally.

On the other hand, we have argued \cite{Hlubina02} that the smallness
of the Josephson product does not follow from the particular
properties of the barrier, but is rather an intrinsic property of the
cuprates.  Such a point of view has been first advocated in
\cite{Deutscher99}.  However, that paper did not consider alternative
more conventional explanations.  In order to support our point of
view, in \cite{Hlubina02} we have discussed the Josephson product for
intrinsic Josephson junctions in the $c$-axis direction.  Such
junctions can be viewed as an analogue of $ab$-plane break junctions
(since the misorientation angle vanishes for both), but are preferable
because of simpler geometry of the interface. Moreover, zero energy
surface bound states which may develop at $ab$-plane surfaces because
of the $d$-wave symmetry of the pairing state \cite{Hu94} do not form
in the $c$-axis direction, simplifying the analysis of intrinsic
Josephson junctions.  By an analysis of the experimental data
\cite{Latyshev99}, we came to the conclusion that the experimental
Josephson product is reduced with respect to theoretical predictions.
Thus we have shown that although the barriers in grain boundaries and
in intrinsic Josephson junctions are of very different nature, both
types of junctions exhibit a suppressed Josephson product. Therefore
we concluded that this suppression is not due to specific barrier
properties as suggested in \cite{Halbritter92}, but rather due to some
intrinsic property of the high-$T_c$ superconductors.


\vspace{1cm}
\noindent
3. EXPERIMENT

\begin{figure}[ht]
\centerline{\includegraphics[width=10cm]{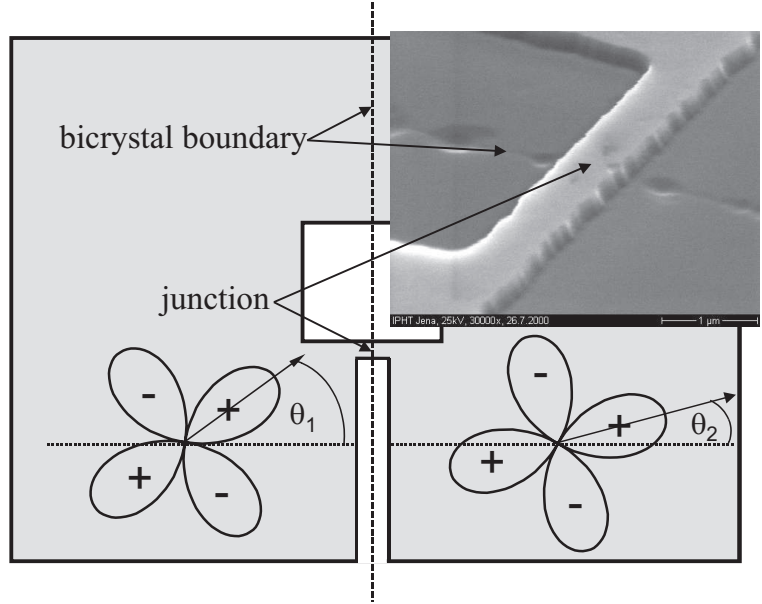}}
\caption{Schematic picture of the RF SQUID (taken from
\cite{Ilichev01}).  The YBCO thin film occupies the gray area.  The
inset shows an electron microscope image of the narrow grain boundary
Josephson junction.}
\label{Fig:Photo}
\end{figure}

\vspace{0.5cm}
\noindent
As shown in the previous Section, theory predicts anomalous behavior
of the Josephson current in the cuprates as a function of the phase
difference $\varphi$ and temperature. One of the most striking
predicted phenomena is a non-monotonic temperature dependence of the
Josephson critical current, forced by the negative contribution of
anomalous Andreev states. However, this has not been observed by
standard means in spite of enormous efforts of many groups. The main
obstacle seems to originate from the interface roughness which leads
to a suppression of the anomalous Andreev states
\cite{Matsumoto95,Barash96}. Even the most promising type of Josephson
junctions, grain boundary junctions, contain defects on a
characteristic scale of 1~$\mu$m (Fig.~2 of
\cite{Mannhart96}) coming from defects in the bicrystal substrate, see
Fig.~\ref{Fig:Photo}. A similar problem is relevant also in $c$-axis
Josephson junctions, where a large second harmonic of the Josephson
current is expected. In that case, in order to avoid tunneling in the
$ab$ plane direction and to achieve a high transparency of the
interface, atomically flat surfaces are required.

Thus, in order to observe the predicted anomalies, one should avoid
the structural defects at the interface. Following this strategy the
$c$-axis Josephson junctions were prepared in situ by covering YBCO by
thick Au layer preventing the degradation of the YBCO surface during
processing. On the other hand, we have chosen the position of the
grain boundary junctions in such a way that they were placed between
the defects of the substrate shown in Fig.~\ref{Fig:Photo}.  Therefore
we were forced to reduce the size of the junction below 1~$\mu$m.
However, for submicron $45^\circ$ junctions where the anomalies should
be most pronounced, the Josephson coupling energy $E_J$ is comparable
with $T$ in the temperature range of interest. This leads to large
phase fluctuations and consequently to a suppression of the Josephson
current. At first sight it seems that the Josephson current of
submicron $45^\circ$ junctions cannot be measured. But this is not so.

\vspace{0.5cm} \noindent 3.1. EXPERIMENTAL METHOD.
Phase fluctuations can be suppressed by connecting the Josephson
junction into a ring (rf SQUID), since in that geometry the phase
difference across the junction is controlled by the large phase
stiffness of the superconducting ring.  This technique therefore
offers a unique possibility to study Josephson junctions at
temperatures $T$ much higher than the junction energy. In addition,
one can change the phase difference $\varphi$ on the Josephson
junction by applying external magnetic flux $\Phi_{dc}$ to the ring
\begin{equation}
\varphi=\varphi_{dc}-\frac{2\pi L_s}{\Phi_0}I(\varphi),
\label{Eq:phi}
\end{equation}
where $\varphi_{dc}=2\pi\Phi_{dc}/\Phi_0$, $L_s$ is the inductance of
the rf SQUID, and $I(\varphi )$ is the Josephson current.  We have
used the modified Rifkin-Deaver method \cite{Rifkin76,Ilichev01a} to
restore the current-phase dependence $I(\varphi)$.  The method is
simple: The rf SQUID is inductively coupled to a high-quality parallel
resonance circuit driven at its resonant frequency $\omega_0$. The
angular phase shift $\alpha$ between the rf driving current $I_{rf}$
and the voltage across the circuit is measured by a rf lock-in
voltmeter as a function of the external magnetic flux. The flux is set
by a dc current $I_{dc}$ through the coil of the resonant circuit with
inductance $L_T$. Thus the total external magnetic flux can be
expressed as $\varphi_e=\varphi_{dc}+\varphi_{rf}$, where
$\varphi_{dc}=2\pi MI_{dc}/\Phi_0$ and $\varphi_{rf}=2\pi M
I_{rf}/\Phi_0$ and $M$ is the mutual inductance between the rf SQUID
and the resonant circuit. The rf SQUID and the resonant circuit are
characterized by quality factors $q=\omega_0L_s/R$ and
$Q=R_T/\omega_0L_T$, respectively. The analysis of such a system can
be considerably simplified in the adiabatic ($q\ll 1$, $Q\gg 1$),
small signal regime ($\varphi_{rf}\ll 1+(2\pi
L_s/\Phi_0)dI(\varphi)/d\varphi$), when the rf SQUID follows
adiabatically the signal from the resonant circuit and the internal
flux can be expressed by Eq.~\ref{Eq:phi}. $I(\varphi_{dc} )$ is
calculated from the experimental $\alpha(\varphi_{dc})$ data using
\begin{equation}
I(\varphi_{dc})=\frac{L_TI_0^2}{2\pi
Q\Phi_0}\int_0^{\varphi_{dc}}\tan\alpha
(\varphi_{dc})d\varphi_{dc}, \label{Eq:Iphi}
\end{equation}
where $I_0$ is the period of $\alpha(I_{dc})$ (i.e.  $MI_0=\Phi_0$)
and the quality factor $Q$ is measured independently from the width of
the resonance curve of the parallel resonant circuit. Using
Eqs.~\ref{Eq:phi},\ref{Eq:Iphi} $I(\varphi)$ can be restored. This
method, being differential with respect to $\varphi$, provides a high
sensitivity of the current phase measurement.

Let us emphasize that the critical current determined by this method
does not depend on the inductance of the rf SQUID. In fact, one can
easily show that $I_c=I(\varphi_{dc}^0)$, where
$\tan\alpha(\varphi^0_{dc})=0$ can be determined from experimental
data by requiring $\int_0^\pi\tan\alpha(\varphi_{dc})
d\varphi_{dc}=0$. The last condition comes from the periodicity of
$I(\varphi)$ and enables a subtraction of a constant phase shift
coming from electronics and cables. Since the only sample
characteristics entering Eq.~\ref{Eq:Iphi} is $I(\varphi)$, the
measurement system can be calibrated using samples with known critical
current and $I(\varphi)$. We have usually carried out the calibration
making use of Nb rf SQUIDs.  Typical values of the quantities used in
our experimental setup are: $L_s=80$~pH, $L_t=0.27\ \mu$H,
$Q=155$. Critical currents measured by the present method and by
standard transport measurements were in good coincidence. As already
explained, $I(\varphi)$ is measurable even if the thermal energy
exceeds the Josephson coupling energy. In fact, critical currents down
to 50 nA were recently detected at $T=4.2$ K \cite{Ilichev01} using
cooled preamplifier placed near the parallel resonant circuit.  All
$I(\varphi)$ measurements presented in this paper have been performed
in a gas-flow cryostat with a five-layer magnetic shielding in the
temperature range $1.6\le T<90$~K.

\begin{figure}[ht]
\centerline{\includegraphics[width=8cm]{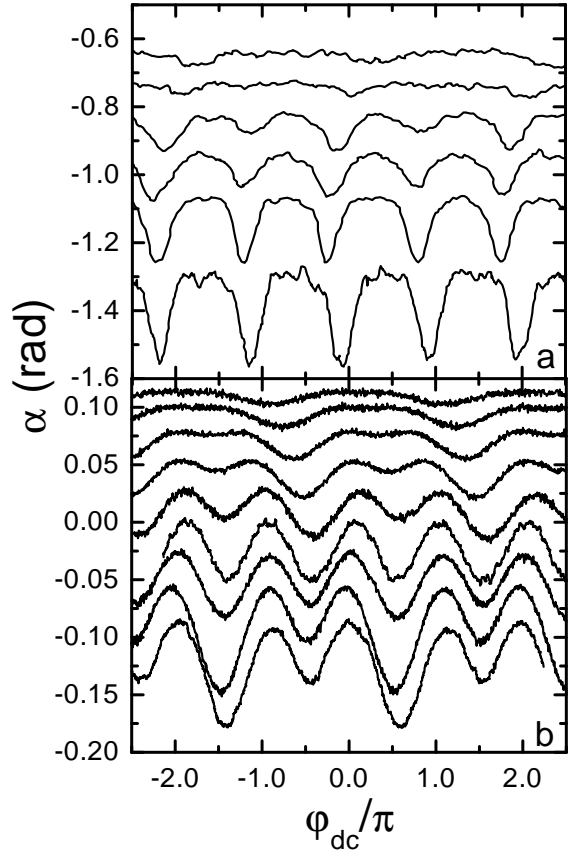}}
\caption{The phase angle $\alpha$ as a function of $\varphi_{dc}$
measured at different temperatures. (a) Asymmetric $45^\circ$ grain
boundary \cite{Ilichev99}. Top to bottom curves correspond to
$T=$40, 30, 20, 15, 10, and 4.2 K, respectively. (b) Symmetric
$45^\circ$ grain boundary \cite{Ilichev01}. Top to bottom curves
correspond to $T=$35, 30, 25, 20, 15, 11, 10, 5, and 1.6 K,
respectively.} \label{Fig:Alpha45}
\end{figure}

\vspace{0.5cm} \noindent 3.2. GRAIN BOUNDARY JOSEPHSON JUNCTIONS.
The rf SQUIDs were prepared at IPHT Jena by laser deposition of YBCO
thin films of thickness 100~nm on bicrystal substrates. They were
patterned in the shape of a square washer 3500~$\mu$m$\times
3500~\mu$m with a hole $50$~$\mu$m$\times 50$ $\mu$m by electron beam
lithography (see Fig.~\ref{Fig:Photo}). The estimated Josephson
penetration depth $\lambda_J$ is much smaller than the width of the
wide junction, $w_l=1725$ $\mu$m, and larger than the width of the
narrow junction, $w_s=1\ \mu$m and $w_s=0.7\ \mu$m for asymmetric and
symmetric $45^\circ$ junctions, respectively. Thus the behavior of the
rf SQUID is dictated by the narrow junction only. For symmetric
$45^\circ$ junctions the submicron bridge was formed at a position
between the defects of the substrate which are visible in
Fig.~\ref{Fig:Photo}. The experimental results are shown in
Fig.~\ref{Fig:Alpha45}. Local minima appear at low temperatures on the
$\alpha (\varphi_{dc})$ curve close to $\varphi_{dc}=2\pi n$ where $n$
is integer. The existence of the local minima dictates
\begin{equation}
\left. \frac{d^3I(\varphi)}{d\varphi^3}\right|_{\varphi=2\pi n}>0.
\label{eq:min}
\end{equation}
Note that neither the conventional tunneling theory nor the I and
II theories by Kulik and Omelyanchuk predict such local minima on
the derivatives of CPR \cite{Likharev79}.

To demonstrate that the anomalous behavior of the Josephson current is
a peculiarity of $45^\circ$ junctions we have measured $I(\varphi)$ of
symmetric YBCO grain boundary junctions with misorientations
$\theta=24^\circ$ and $36^\circ$. The results of this study are shown
in Fig.~\ref{Fig:Comp}. We have found that the critical current
decreases exponentially with the misorientation angle $\theta$,
$\exp(-\theta/\theta_0)$ with $\theta=5.6^\circ$, in good agreement with
\cite{Hilgenkamp98}. Moreover, both for $\theta=24^\circ$ and $36^\circ$,
no considerable deviation from sinusoidal dependence of $I(\varphi)$
was observed, implying tunneling as the dominant transport
mechanism.

\begin{figure}[ht]
\centerline{\includegraphics[width=8cm]{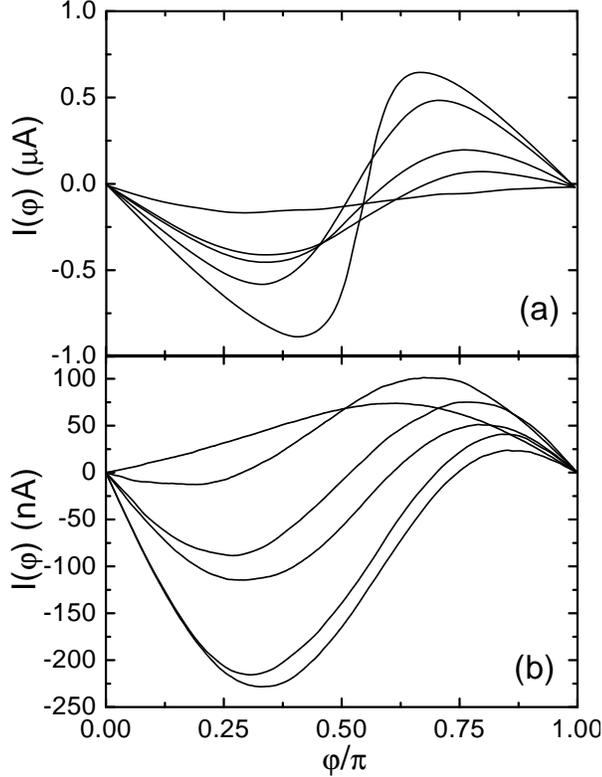}}
\caption{Current-phase relation for asymmetric (a) and symmetric (b)
$45^\circ$ grain boundary junctions (taken from \cite{Grajcar02}). (a) Top to
bottom curves at $\varphi=0.75\pi$: $T=4.2$, 10, 15, 20, and 30~K. (b)
Bottom to top curves at $\varphi=0.375\pi$: $T=1.6$, 5, 10, 11, 15,
and 25~K.} \label{Fig:Ip45}
\end{figure}

\begin{figure}[ht]
\centerline{\includegraphics[width=8cm]{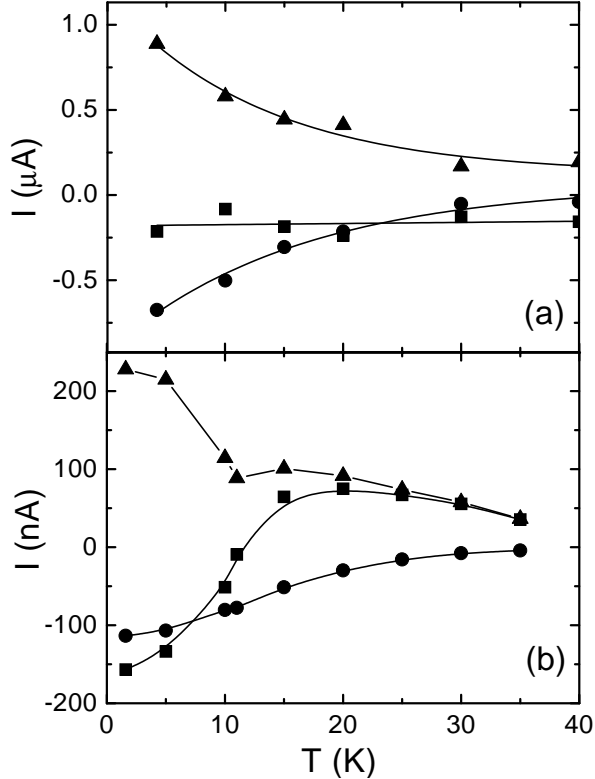}}
\caption{Critical current $I_c$ (triangles) and the first two
harmonics $I_1$ (squares) and $I_2$ (circles) as a function of
temperature for asymmetric (a) and symmetric (b) grain boundary
$45^\circ$ junctions (taken from \cite{Grajcar02}). The lines are
guides to the eye.}
\label{Fig:I1I2}
\end{figure}

Let us return to the case of $45^\circ$ junctions now. Theory predicts
that the first harmonic should be suppressed for such junctions.  If
we in addition take into account that the barrier transmission is low,
we can model $I(\varphi)$ of such junctions by neglecting terms of
higher orders ($n>2$) in Eq.~\ref{eq:Fourier}. Therefore the condition
(\ref{eq:min}) for the existence of the local minima at
$\varphi_{dc}=2\pi n$ dictates $I_2/I_1 < -1/8$.  Thus we conclude
that the $45^\circ$ junctions exhibit anomalously large second
harmonic of the Josephson current which has opposite sign in
comparison with $I_1$. The current-phase relations calculated from the
experimental data in Fig.~\ref{Fig:Alpha45} are shown in
Fig.~\ref{Fig:Ip45}. Note the anomalous form of $I(\varphi)$ at low
temperatures. The temperature dependence of the first two harmonics
$I_1$ and $I_2$, determined from a Fourier analysis of $I(\varphi)$,
is shown in Fig.~\ref{Fig:I1I2}. For both asymmetric and symmetric
$45^\circ$ junctions the second harmonic monotonically increases as the
temperature decreases. The first harmonic is more or less constant for
asymmetric junctions whereas for symmetric $45^\circ$ junctions the
first harmonic starts to exhibit a downturn below 20~K. The most
striking result is that for $T$=12~K, $I_1$ changes sign. In the same
temperature region where $I_1$ starts to exhibit a downturn, the
absolute value of $I_2$ rises from a negligible value at high
temperatures to values comparable to $I_1$ at low temperature. This
experimental fact suggests a common origin of both phenomena. Similar
effects were theoretically predicted for Josephson junctions between
$d$-wave superconductors (Section~2) and our experiments can be
therefore regarded as an independent experimental test of the $d$-wave
symmetry of pairing in YBCO.

\vspace{0.5cm} 
\noindent 
3.2. $c$-AXIS JOSEPHSON JUNCTIONS. The $c$-axis Josephson junctions
were fabricated and characterized jointly by the IRE Moscow and
Chalmers groups following \cite{Komissinski99}. Epitaxial
(001)-oriented YBCO thin films with thickness $150$ nm were obtained
by laser deposition on (100) LaAlO$_{3}$ and (100) SrTiO$_{3}$
substrates and {\it in situ} covered by a $8\div 20$ nm thick Au
layer, thus preventing the degradation of the YBCO surface during
processing. Afterwards, $200$~nm thick Nb counterelectrodes were
deposited by DC-magnetron sputtering.  Junctions with dimensions
$10\times 10\ \mu$m$^{2}$ were formed by photolithography and low
energy ion milling techniques.  The interface resistance per unit area
$R_\Box=R_NS$ (where $R_N$ is the normal state resistance and $S$ is
the junction area) was $R_\Box=10^{-5}\div 10^{-6}\
\Omega\cdot$cm$^{2}$.  Details of the junction fabrication were
reported elsewhere \cite{Komissinski99}.

Surface quality of the YBCO films is very important when current
transport in the $c$-direction is investigated. High-resolution
atomic force microscopy  reveals a smooth surface consisting of
approximately 100~nm long islands with vertical peak-to-valley
distance of $3\div 4$~nm \cite{Komissinski02}. We can exclude that
substantial $ab$-plane tunnel currents flow between YBCO and Nb at
the boundaries of these islands. In fact, theory predicts
formation of midgap states at the surface of semi-infinite CuO$_2$
planes \cite{Hu94,Tanaka95}. Therefore zero bias conductance peaks
should be expected in the $I$-$V$ characteristics at temperatures
larger than the critical temperature of Nb, if the contribution of
$ab$-plane tunneling was nonnegligible.  However, no such peaks
have been observed for all fabricated Nb/Au/YBCO junctions.
Moreover, from the size of the islands and from the vertical
peak-to-valley distance we estimate that the area across which
$ab$-plane tunneling might take place is only $\approx 6$\% of the
total junction area.  However, since the interface resistances per
square are of the same order of magnitude \cite{Sun96} for both,
the $c$-axis and $ab$-plane junctions, we conclude that $ab$-plane
tunneling from YBCO, if present, is negligibly small.

More than 20 junctions were characterized by transport
measurements. At small voltages typical $I$-$V$ curves can be
described by the resistively shunted junction model with a small
capacitance \cite{Likharev79}. Typical critical current densities were
$j_{c}=1\div 12$~A/cm$^2$ and $j_{c}R_\Box=10\div 90\ \mu$V.  The
differential resistance $vs.$ voltage dependence $R_{d}(V)$ exhibits a
gap-like structure at $V\approx 1.2$~mV. This structure has a BCS-like
temperature dependence and disappears at $T_{cR}\approx 9.1$~K,
therefore we ascribe it to the superconducting energy gap of Nb.

\begin{figure}[ht]
\centerline{\includegraphics[width=10cm]{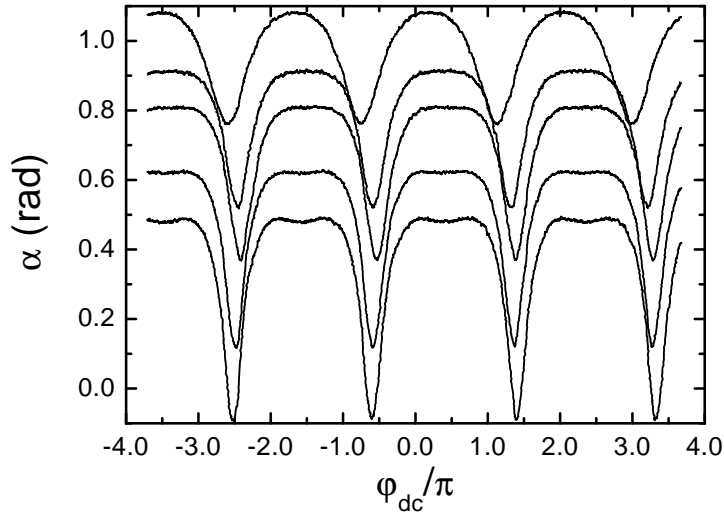}}
\caption{Phase shift $\alpha$ as a function of $\varphi_{dc}$ for
a $c$-axis junction at $T$=1.7, 2.5, 3.5, 4.2, and 6.0~K (from
bottom to top). Taken from \cite{Komissinski02}.} 
\label{Fig:Alphac}
\end{figure}

\begin{figure}[ht]
\centerline{\includegraphics[width=10cm]{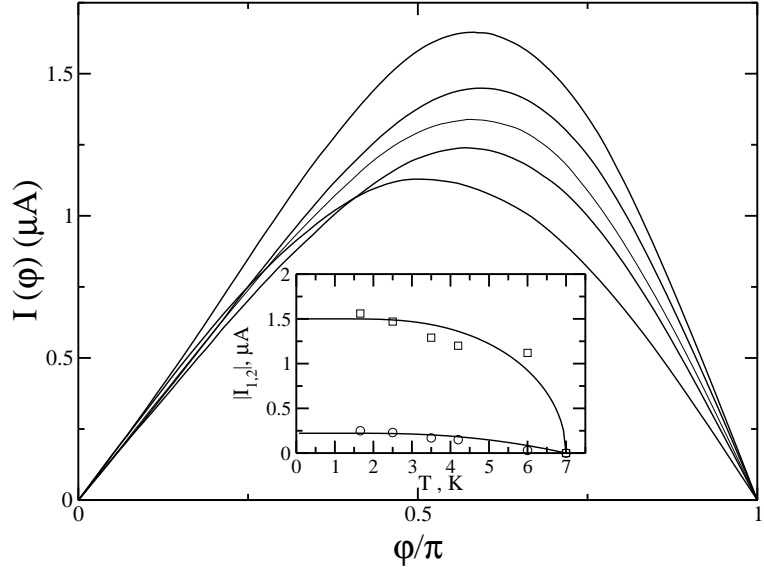}}
\caption{The current-phase relation $I(\varphi )$ of the $c$-axis
junction from Fig.~\ref{Fig:Alphac} at $T$=1.7, 2.5, 3.5, 4.2, and
6.0~K (from top to bottom).  Inset: Temperature dependence of $I_1$
(squares) and $|I_2|$ (circles). Solid lines are fits to
Eqs.~(\ref{eq:j_1},\ref{eq:j_2}) using
$\Delta_R(T)=\Delta_R(0)\tanh[\Delta_R(T)T_{cR}/\Delta_R(0)T]$.
Taken from \cite{Komissinski02}.}
\label{Fig:Ipc}
\end{figure}

The current phase relation of the $c$-axis Josephson junctions was
measured by closing the Nb/Au/(001)YBCO heterostructure into a
superconducting ring with the same geometry as for $45^\circ$ grain
boundary junctions. The experimental results are shown in
Figs.~\ref{Fig:Alphac},\ref{Fig:Ipc}. When compared with $45^\circ$
junctions, the second harmonic of the Josephson current was
considerably smaller but still anomalously large, leading to local
minima of $\alpha(\varphi_{dc})$. As follows from the analysis in
Section~2.2, the opposite signs of the first and second harmonics of
the Josephson current provide direct evidence that in our YBCO samples,
pairing with a macroscopic $d+s$ symmetry is realized.  The large
first harmonic has to be due to an uncompensated $s$-wave component
whose origin is discussed in detail in the next Section.

\vspace{1cm}
\noindent
4. DISCUSSION

\vspace{0.5cm}
\noindent
4.1. GRAIN BOUNDARY JUNCTIONS.  {\it Asymmetric $45^\circ$ junctions.}
The large second harmonic observed in \cite{Ilichev99} confirms the
naive expectations based on the symmetry of the pairing state in the
cuprates. However, at the time of writing the paper \cite{Ilichev99},
we were not able to explain the details of the shape of the
current-phase relation. In particular, we could not find any mechanism
leading to a current-phase relation which was steep in the minima and
flat in the maxima of energy. For instance the Kulik-Omelyanchuk
theory \cite{Likharev79} leads to an exactly opposite picture of jumps
in $I(\varphi)$, which occur in the maxima of energy (at
$\varphi=\pm\pi$). This is a consequence of the fact that for
$\varphi=\pm\pi$ two locally stable branches of the junction energy
cross, one centered around $\varphi=0$ and another one around
$\varphi=\pm 2\pi$.  We believe that in the present work we have found
one physically plausible scenario for how the pattern of jumps of
$I(\varphi)$ observed in \cite{Ilichev99} can be explained. In fact,
Fig.~\ref{Fig:Ipsteep} shows that behavior qualitatively similar to
the results of \cite{Ilichev99} obtains if the Meissner energy
dominates the junction energy Eq.~\ref{eq:F_asym}.  A more detailed
investigation of this idea is under way.  A preliminary analysis
indicates that in order to be applicable to \cite{Ilichev99}, also the
effect of a spatially fluctuating barrier similar to that discussed at
the end of Section 4.2 needs to be taken into account.

Although consistent with naive expectations, under closer inspection,
the experimentally found large value of the second harmonic remains
mysterious:

In the flat scenario, the second harmonic is given by
Eq.~\ref{eq:iphi_asym}.  On the other hand, by combining
Eqs.~\ref{eq:Walker_rough},\ref{eq:Jos_prod} we estimate that the
first harmonic $j_1R_\Box\approx\pi\alpha^2\delta\Delta/e$, where
$\delta$ is the deviation of the misorientation angle $\theta$ from
the nominal value of $45^\circ$.  Taking a realistic value of such a
deviation at least $\delta \sim 10^{-2}$, by comparing to
Eq.~\ref{eq:iphi_asym} we find that the experimental value
$j_2/j_1\sim 1$ can be explained, only if $\vartheta_0\sim 1$ (and
therefore the tunneling has to be possible in a wide range of impact
angles). In this estimate we have made use of the well established
\cite{Hilgenkamp02} value ${\cal D}\sim 10^{-4}$ and we have
taken $\Delta/T\sim 10^2$ at helium temperatures. Note that in order
to obtain $j_2/j_1\sim 1$ we had to assume that only the first
harmonic is renormalized by $\alpha^2\sim 10^{-1}$, while the second
harmonic has to remain unrenormalized.  This implies very peculiar
microscopic physics.

Now let us consider the faceted scenario.  As an order of magnitude
estimate, we use the result Eq.~\ref{eq:j_millis} for the second
harmonic. Unlike in the $c$-axis case, in the present case all
penetration depths have to be set equal to the in-plane penetration
depth of the cuprates $\lambda\approx 0.15 \mu$m. This yields
$j_2=(8\pi\Phi_0)^{-1}\mu_0j_0^2a^2\lambda_1$, where
$\lambda_1=2\lambda\sqrt{1+2(\pi\lambda/a)^2}$. For $a$ we take the
typical length scale of faceting, $a\approx 0.1 \mu$m, since the twin
size is even smaller.  This yields $\lambda_1\approx 2 \mu$m.  The
only unknown in the expression for $j_2$ is the local current density
$j_0$. In what follows we determine $j_0$ from $j_0=\sqrt{8\pi
j_2\Phi_0/(\mu_0 a^2 \lambda_1)}$ where for $j_2$ we take the
experimental critical current density $j_2\sim 10^4$~A/cm$^2$
\cite{Hilgenkamp02}, and we obtain a reasonable value 
$j_0\sim 10^6$~A/cm$^2$. Note that since $B_{\rm
eff}=(2\pi)^{-1}\mu_0j_0a\approx 3\times 10^{-4}$~T and
$\Phi_0/(2a\lambda_1)\approx 5\times 10^{-3}$~T, the criterion
Eq.~\ref{eq:crit_millis} for the applicability of
Eq.~\ref{eq:j_millis} is well satified.  On the other hand, we
estimate the first harmonic from ${j_1/j_0}\approx (a/L)^{1/2}$, which
is a random walk-type formula, indicating that $j_1$ averages to zero
in a sufficiently long junction. The experiment requires $j_2>j_1$.
This is possible only for sufficiently long junctions, $L\sim 1$ mm,
whereas in \cite{Ilichev99} much shorter junctions with $L\sim 1\mu$m
were studied. Therefore within the faceted scenario it is not possible
to explain the large measured $I_2/I_1$ ratio.

We conclude that both within the flat and faceted scenaria, the second
harmonic measured in \cite{Ilichev99} appears to be anomalously large.

{\it Symmetric $45^\circ$ junctions.}  The anomalous temperature
dependence of the first harmonic and the large second harmonic
observed in this type of junctions \cite{Ilichev01} is qualitatively
consistent with the theoretical prediction Eq.~\ref{eq:iphi_sym2} and
not with Eq.~\ref{eq:iphi_sym3}.  The data seems to imply (as was the
case for asymmetric junctions as well) that the impact angle
dependence of the barrier transmission is weak.

In what follows we attempt a more quantitative discussion of the
results found in \cite{Ilichev01}.  According to Fig.~\ref{Fig:I1I2}
the first harmonic changes sign at a temperature $T^\ast\approx
12$~K. This together with the estimate $\Delta\approx 20$~meV
\cite{Alff98} requires $D(\pi/4)/D(0)\approx 0.1$, i.e. the barrier,
although presumably quite thin, can't be modelled by a delta function.

Now let us compare the relative magnitudes of the first and second
harmonics. Theory predicts that for $T>T^\ast$, the maximal value of
$j_1$ is $j_1^{\rm max}/j_L\approx D(0)$.  On the other hand, from
Eq.~\ref{eq:iphi_sym2} it follows that at $T=T^\ast$ the second
harmonic $|j_2(T^\ast)|/j_L\approx D(0)^3/24D(\pi/4)$.  Therefore
according to theory $|j_2(T^\ast)|/j_1^{\rm max}\approx
D(0)^2/24D(\pi/4)\approx D(0)/2$, whereas from Fig.~\ref{Fig:I1I2} we
estimate $|j_2(T^\ast)|/j_1^{\rm max}\approx 1$.  Thus theory can
describe the experimental results only if $D(0)\sim 1$.

However, in what follows we show that $D(0)\ll 1$ and therefore the
experimental second harmonic is again too large when compared with
simple minded theory. In fact, in order that higher-order harmonics
are negligible \cite{Ilichev01} even at the lowest studied
temperatures $T_{\rm min}\approx$~1.6~K, we require that the typical
mid-gap state energy $\Delta\sqrt{D(\pi/4)/2}<2T_{\rm min}$, yielding
$D(\pi/4)<4\times 10^{-4}$, and therefore $D(0)\approx 10
D(\pi/4)<4\times 10^{-3}$.  Moreover, $D(0)\sim 10^{-3}$ seems to be
consistent with the exponential decrease of $j_c$ with the
misorientation angle $\theta$ \cite{Hilgenkamp98}.

In \cite{Grajcar02} we have proposed that the first harmonic might by
suppressed by interface roughness according to
Eq.~\ref{eq:Walker_rough}. This would require a very small roughness
parameter $x\sim 10^{-3}$ and therefore the junction would have to be
nearly completely rough. However, we have overlooked the fact that for
such junctions it is hard to believe that the deviation $\delta$ from
the misorientation angle $\theta=45^\circ$ can be sufficiently small
in order to keep the first term in Eq.~\ref{eq:Walker_rough} small.

In summary, the second harmonic of $45^\circ$ grain boundary junctions
seems to be too large to be explicable by conventional theory.

\vspace{0.5cm}
\noindent
4.2. $c$-AXIS JUNCTIONS. Let us start by estimating the transparency of
the barrier between YBCO and Nb from the normal-state resistance per
unit area $R_\Box$. According to the band-structure calculations (for a
review, see \cite{Pickett89}), the hole Fermi surface of YBCO is a
slightly warped barrel with an approximately circular in-plane
cross-section (to be called Fermi line) with radius $k_F$.  In what
follows, we represent the electron wavevector $\mathbf{k}$ in
cylindrical coordinates, $\mathbf{k}=(k,\theta,k_z)$. We estimate the
uncertainty of the in-plane momentum as $\delta k\approx 2\pi/l$,
where $l$ is the characteristic size of the islands on the YBCO
surface (fig.1). We evaluate $R_\Box$ making use of the Landauer formula
and note that only tunneling from a shell around the Fermi line with
width $\delta k$ is kinematically allowed. The barrier transparency
$D(\theta)$ depends on the details of the $c$-axis charge dynamics in
YBCO, with maxima in those directions $\theta$, in which the YBCO
$c$-axis Fermi velocity $w(\theta)$ is maximal. Since for
$\theta=\pi/4$ and symmetry equivalent directions $w(\theta)$ is
minimal \cite{Xiang96}, we expect that there will be 8 maxima of
$D(\theta)$ on the YBCO Fermi line where $D(\theta)\approx D$, which
are situated at $\theta=\theta_0$ and symmetry equivalent
directions. The modulation of the function $D(\theta)$ depends on the
thickness of the barrier between YBCO and Nb \cite{Wolf85}. We
consider two limiting distributions of the barrier transparency
$D(\theta)$ along the YBCO Fermi line: (a) a featureless
$D(\theta)\approx D$ and (b) a strongly peaked $D(\theta)$, roughly
corresponding to thin and thick barriers, respectively
\cite{Wolf85}. In the thick barrier limit the angular size of the
maxima of $D(\theta)$ can be estimated as $\delta\theta\approx \delta
k/k_F$. With these assumptions we find
\begin{equation}
R_\Box^{-1}={\frac{\langle D\rangle e}{\Phi_0}}A,  
\label{eq:conductivity}
\end{equation}
where $A$ measures the number of conduction channels and
$\langle\ldots\rangle$ denotes an average over the junction area. In
the thin and thick barrier limits, we find $A\approx k_F\delta k/\pi$
and $A\approx 2\delta k^2/\pi$, respectively. Taking $l\approx 100$ nm
and $k_F\approx 0.6$ \AA$^{-1}$ \cite{Shen95}, the measured
$R_\Box=6\times 10^{-5}$ $\Omega$cm$^2$ can be fitted with $\langle
D\rangle_{\mathrm{thin} }\approx 1.7\times 10^{-5}$ and $\langle
D\rangle_{\mathrm{thick}}\approx 8.3\times 10^{-4}$.

Now we can turn to the discussion of $I(\varphi)$.  Since we have
observed no midgap surface states in the $R_d(V)$ curves, we can
neglect the surface roughness, and the Josephson current can be
calculated from \cite{Zaitsev84}
\begin{equation}
I(\varphi)={\frac{2e}{\hbar}}\sum_{k,\theta} k_BT\sum_{\omega} {\frac{
D\Delta_R\Delta_{\mathbf{k}}\sin\varphi}{2\Omega_R\Omega_{\mathbf{k}}+D\left[
\omega^2+\Omega_R\Omega_{\mathbf{k}} +\Delta_R\Delta_{\mathbf{k}}\cos\varphi
\right]}},
\end{equation}
where the sum over $k,\theta$ is taken over the same regions with
areas $A$ as in Eq.~(\ref{eq:conductivity}), $\Delta_R$ and
$\Delta_{\mathbf{k}}$ are the Nb and YBCO gaps, respectively, and
$\Omega_i=\sqrt{\omega^2+\Delta_i^2}$ with $i=R,\mathbf{k}$. Keeping
only terms up to second order in the (small) junction transparency
$D$, the first and second harmonics of the Josephson current densities
for an untwinned YBCO sample read
\begin{eqnarray}
j_0(T)R_\Box&\approx&{\frac{\Delta_s}{\Delta_d^\ast}}{\frac{\Delta_R(T)}{e}},
\label{eq:j_1} \\
j_2(T)R_\Box&\approx&-{\frac{\pi}{8}} {\frac{\langle D^2\rangle}{\langle
D\rangle}} {\frac{\Delta_R(T)}{e}}\tanh\left({\frac{\Delta_R(T)}{2k_BT}}
\right),  
\label{eq:j_2}
\end{eqnarray}
where $\Delta_d^\ast=\pi\Delta_d[2\ln(3.56\Delta_d/T_{cR})]^{-1}$ and
$ \Delta_d^\ast=\Delta_d|\cos 2\theta_0|$ in the thin and thick
barrier limits, respectively. In Eqs.~(\ref{eq:j_1},\ref{eq:j_2}) we
have assumed that the angular variation of the YBCO gap can be
described as $\Delta(\theta)=\Delta_s+\Delta_d\cos 2\theta$, where
$\Delta_d$ and $\Delta_s$ are the $d$-wave and $s$-wave gaps.  We have
assumed that $\Delta_d^\ast$ is larger than both, $\Delta_R$ and
$\Delta_s$. The factor $\Delta_s/\Delta_d^\ast$ can be estimated from
the measured $j_0R_\Box$ products for Josephson junctions between
untwinned YBCO single crystals and Pb counterelectrodes. For such
junctions $j_0(0)R_\Box\approx 0.5\div 1.6$~mV \cite{Sun96}. Using the
Pb gap $\Delta_R=1.4$~meV in Eq.~(\ref{eq:j_1}), we obtain
$\Delta_s/\Delta_d^\ast\approx 0.36\div 1.1$.

From the relative sign of $I_{1}$ and $I_{2}$ we know that the finite
first harmonic has to be due to the macroscopic $d+s$ symmetry of our
YBCO sample. The simplest way how this can be realized is to assume
that the numbers of the two types of twins are unequal.  In fact,
detailed structural studies show that this is the case for
sufficiently thin YBCO films even if they are grown on the cubic
substrate SrTiO$_{3}$\cite{Didier97}. If we denote the twin fractions
as $(1+\delta )/2$ and $(1-\delta )/2$, then the measured first
harmonic of the CPR, $j_{1}$, is proportional to the deviation from
equal population of twins, $j_{1}=\delta j_{0}$
\cite{ODonovan97}. Using $\Delta _{R}=1.2$~meV determined from the
$R_{d}(V)$ data and our estimate $\Delta _{s}/\Delta _{d}^{\ast
}\approx 0.36\div 1.1$, we find that the measured first harmonic
$j_{1}$ can be fitted with $\delta \approx 0.07\div 0.21$, which is in
qualitative agreement with
\cite{Didier97}, where $\delta \approx 0.14$ for 1000 \AA\ thick YBCO
films has been observed.

Fitting the measured $j_2R_\Box$ by Eq.~(\ref{eq:j_2}), $\langle
D^2\rangle/\langle D\rangle\approx 3.2\times 10^{-2}$ was found, which
is much larger than both $\langle D\rangle_{\mathrm{thin}}$ and
$\langle D\rangle_{\mathrm{thick}}$. Similarly as in the case of grain
boundary junctions, this means that the experimental second harmonic
is too large when compared with naive estimates. In what follows we
discuss the possible causes of such behavior.

Let us start by considering the faceted scenario. From the known
values of $j_0R_\Box$ \cite{Sun96} and $R_\Box$ we estimate
$j_0\approx 8\div 27$~A/cm$^2$. Taking for a typical twin size
$a\approx 10$~nm, the effective magnetic field from Section 2.3 is
estimated as $B_{\rm eff}\approx (1.6\div 5.4)\times
10^{-10}$~T. Moreover, since $\lambda_R\approx 39$~nm, $\lambda\approx
240$~nm, and $\lambda_c\approx 3$~$\mu$m (as an upper bound, for the
cuprates we take numbers which are valid for underdoped YBCO
\cite{Cooper94}), we find $\lambda_1\approx 320$~$\mu$m.  Since this
implies $\Phi_0/(2a\lambda_1)\approx 3\times 10^{-4}$~T, the criterion
Eq.~\ref{eq:crit_millis} is seen to be well satisfied and therefore
the second harmonic can be calculated from Eq.~\ref{eq:j_millis},
yielding $|j_2|R_\Box\approx (3\div 35)\times 10^{-11}$~V, orders of
magnitude smaller than the experimental value $|j_2|R_\Box\approx
15$~$\mu$V.

Thus we were led to look for alternative explanations of the large
second harmonic. In \cite{Komissinski02} we came up with an
explanation which assumed that the junction transparency $D$ is a
fluctuating function of the position $\mathbf{r}$. Adopting the WKB
description of tunneling \cite{Wolf85}, we assumed that the local
barrier transparency $D(s(\mathbf{r}))=\exp(-s_0-s(\mathbf{r}))$,
where $s_0$ is the WKB tunneling exponent and $s(\mathbf{r})$ its
local deviation from the mean.  Assuming a Gaussian distribution of
$s$ with a mean deviation $\eta$, $P(s)\propto\exp(-s^2/\eta^2)$, we
estimated the spatial averages as $\langle
D^n\rangle=\int_{-s_0}^{s_0} ds P(s) D^n(s)$.  In the thin barrier
limit, the values $\langle D^2\rangle_{\mathrm{thin}}=8.6\times
10^{-7}$ and $\langle D\rangle_{\mathrm{thin}}=2.3\times 10^{-5}$
required to fit the experiments correspond to an average WKB exponent
$s_0^{\mathrm{thin}}\approx 15.5$ with $\eta_{\mathrm{thin}}\approx
4.3$. In the thick barrier limit we obtain
$s_0^{\mathrm{thick}}\approx 9.1$ and $\eta_{\mathrm{thick}}\approx
2.8$.

\vspace{0.5cm}
\noindent
4.3. MICROSCOPIC IMPLICATIONS. Very recently it has been pointed out
by one of us \cite{Hlubina02} that the two apparently unrelated
experimental facts, namely the suppressed Josephson product $I_1R_N$
and the enhanced ratio $|I_2/I_1|$, can be explained by a single
assumption that in the cuprates some mechanism is operative which
leads to a suppression of $I_1$, while leaving $R_N$ and $I_2$
intact. In what follows we describe one such mechanism which we
believe to be the most promising one.  Namely, we suggest that at low
temperatures the superconducting state of the cuprates supports
fluctuations of the superconducting phase.  Such fluctuations
presumably do not affect $R_N$, while they do influence the Josephson
current. In simplest terms, if we denote the phases of the
superconducting grains forming the junction as $\phi_i$, then the
fluctuations renormalize the first and second harmonics by the factors
$\langle e^{i\phi_1}\rangle\langle e^{i\phi_2}\rangle$ and $\langle
e^{2i\phi_1}\rangle\langle e^{2i\phi_2}\rangle$, respectively, where
$\langle\ldots\rangle$ denotes a ground-state expectation value.  Thus
experiment requires that the fluctuations have to be of such type that
$|\langle e^{i\phi}\rangle|=\alpha\approx 0.3$ and $|\langle
e^{2i\phi}\rangle|\approx 1$.  Precisely this behavior is expected if
the $d$-wave order parameter fluctuates towards $s$-wave pairing
(which pairing is expected to be locally stable within several
microscopic models of the cuprates).

An independent check of our picture is provided by measurements of the
Josephson product for junctions between the cuprates and low-$T_c$
superconductors. In such a case, we predict that the Josephson product
should be renormalized by a factor $\alpha$ instead of $\alpha^2$ for
junctions between two cuprates. Junctions of this type have been
studied extensively in the past. In particular, $ab$-plane junctions
between YBCO and Pb exhibit Josephson products of $0.2\div 1.2$~mV
\cite{Sun96}. On the other hand, Eq.~\ref{eq:ambegaokar} predicts 
$I_cR_N\approx e^{-1}\Delta_{\rm Pb}\ln(4\Delta_{\rm YBCO}/\Delta_{\rm
Pb})\approx 5.7$~mV in this case, if we assume a (100) YBCO surface
and take $\Delta_{\rm YBCO}\approx 20$~meV and $\Delta_{\rm Pb}\approx
1.4$~meV.  We interpret the large experimental scatter of $I_cR_N$ as
being due to varying $ab$-plane tunneling direction. Thus theory has
to be compared with the largest experimental value, and the
theoretical result has to be multiplied by a renormalization factor
$\alpha\approx 0.2$ in order to bring it in agreement with experiment.
This value is in semiquantitative agreement with $\alpha\approx 0.3$
which was determined from the Josephson product of grain boundary
junctions.  It is worth mentioning that the phase fluctuation picture
also may be relevant for the experiment \cite{Komissinski02}, where a
large second harmonic has been found in a $c$-axis Josephson junction
between YBCO and Nb.


\vspace{1cm}
\noindent
5. CONCLUSIONS

\vspace{0.5cm}
\noindent
In this paper we have tried to argue that the Josephson effect can
provide nontrivial information not only about the symmetry of the
pairing state in the cuprates, but also about the fluctuations of the
superconducting order paramater. The latter are expected to be quite
large especially in the underdoped region, where charge fluctuations
should be suppressed.  A preliminary analysis of the Josephson product
and of the current-phase relation in grain boundary and $c$-axis
junctions indicates \cite{Hlubina02} that the phase fluctuations of
the superconducting order parameter are quite large and of a very
special type, favoring local fluctuations from $d$-wave towards
extended-$s$ pairing. 

Surprisingly, the potential of the Josephson effect as a test of the
phase rigidity of the cuprates seems to have been missed in the past
and therefore quite few studies have attempted quantitative analysis
of the Josephson phenomena.  Therefore there are still more questions
than answers in this field. Some of the most pressing problems of the
field (from our point of view) are listed below.

1. The nature of the barrier in both grain boundary and $c$-axis
junctions is unknown. In the better studied case of grain boundary
junctions, we believe that the barrier is in the tunnel limit and for
sufficiently large misorientation angles there are no pinholes in it.
However, the typical barrier width and height don't seem to be well
known. Therefore also the impact angle dependence of the barrier
transmission is not known a priori. On the other hand, in order to
explain the current-phase relation of $45^\circ$ grain boundary
junctions, we had to assume a weak impact angle dependence. It remains
to be seen whether this agrees with the barrier properties determined
independently, e.g. making use of the STM microscopy where the
junction is viewed from above along a path crossing the grain
boundary. With such a method, both the barrier height and width could
be measurable.

\begin{figure}[ht]
\centerline{\includegraphics[width=10cm]{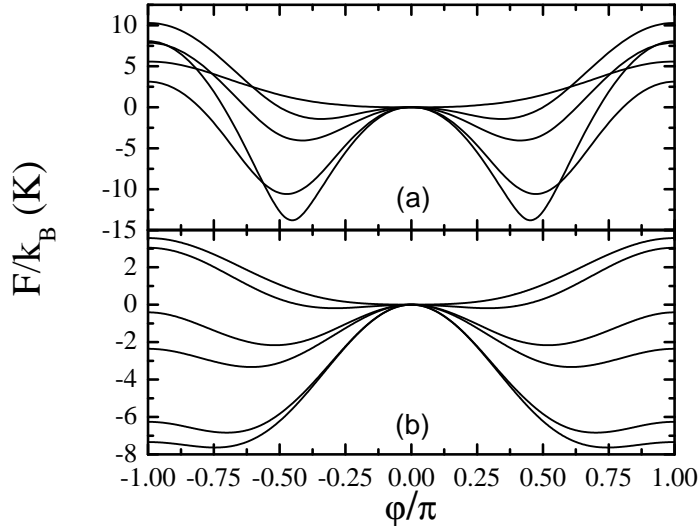}}
\caption{Free energy $F(\varphi)$ as a function of the phase
difference across the weak link (taken from \cite{Ilichev01}).  The
zero of energy has been set so that $F(0)=0$.  (a) Asymmetric
$45^\circ$ grain boundary. Top to bottom curves correspond to $T=$ 40,
30, 20, 15, 10, and 4.2 K, respectively.  (b) Symmetric $45^\circ$
grain boundary. Top to bottom curves correspond to $T=$20, 15, 11, 10,
5, and 1.6 K, respectively.}
\label{Fig:Ephi}
\end{figure}

2. The role of various types of disorder in the junction region
(interface roughness and faceting, spatially fluctuating barrier
height and width, etc.)  has to be studied systematically. In
particular, disorder is believed to reduce the ability of $45^\circ$
grain boundary junctions to support mid-gap states
\cite{Matsumoto95} and therefore diminishes the second harmonic as well. 
On the other hand, simple symmetry arguments predict a suppression of
the first harmonic of $I(\varphi)$ by the surface roughness.  Also an
explicit calculation of the first harmonic in the presence of a finite
barrier roughness supports this conclusion \cite{Barash96}.  A
reliable answer to the question about which of the above two effects
of disorder dominates is therefore crucial in order to decide whether
disorder increases the ratio of the second and first harmonics of the
current-phase relation $|I_2/I_1|$ as suggested in Section~4, or it
rather diminishes it.  If the latter alternative is realized, then the
experimental observation of a large second harmonic provides even
stronger argument in favour of anomalous quantum phase fluctuations in
the (bulk) cuprates.  

3. Interface roughness and disorder in the barrier region may be also
responsible for the weak impact angle dependence of the barrier
transmission, which is implied by the large second harmonics observed
in $45^\circ$ grain boundary junctions.

4. From the point of view of applications, especially the grain
boundary Josephson junctions have attracted a lot of interest
recently. In particular, junctions with sufficiently large second
harmonics support doubly degenerate ground states (see
Fig.~\ref{Fig:Ephi}) and it has been suggested \cite{Ioffe99} that
this property might be exploited in the construction of a `quiet
qubit', i.e. of a two level system which couples only weakly to its
environment and can be used as a basic element in a quantum
computer. This is a fascinating proposal and a lot of efforts is being
spent on its realization. Several nontrivial problems have to be
solved before the final goal can be reached: on the technological
side, a technique for a reproducible fabrication of well behaved
submicron grain boundary junctions has to be developed.  From the
point of view of basic science, new routes to reducing the coupling of
the junction to the environment are to be looked for, in order to make
the junctions really quiet.


\vspace{1cm}
\noindent
ACKNOWLEDGEMENTS

\vspace{0.5cm} \noindent We thank M.~H.~S.~Amin, A.~Golubov,
H.~E.~Hoenig, R.~P.~J.~IJsselsteijn, Z. Ivanov, S. Kashiwaya, P.
V. Komissinski, M.~Yu.~Kupriyanov, S.~A.~Kovtonyuk, H.-G.~Meyer,
A.~N.~Omelyanchouk, G.~A.~Ovsyannikov, V.~Schultze, Y.~Tanaka,
N.~Yoshida, A.~M. Zagoskin, and V.~Zakosarenko for collaborations
on solving the problems discussed in this paper.  We learned a lot
also from discussions with Yu.~S.~Barash, V.~Bez\'ak,
M.~V.~Fistul, H.~Hilgenkamp, R.~Kleiner, J.~Mannhart, R.~G.~Mints,
A.~Plecenik, N.~Schopohl, M.~Sigrist, and A.~Y.~Tzalenchuk. R.~H.
and M.~G. were supported by the Slovak Scientific Grant Agency
under Grant No.~VEGA-1/9177/02 and by the Slovak Science and
Technology Assistance Agency under Grant No.~APVT-51-021602. E.~I.
was supported by DFG (Ho461/3-1). The support by D-Wave Systems is
also acknowledged.

\end{document}